\documentclass[prb,aps,twocolumn,amsmath,amssymb]{revtex4-2}
\usepackage{graphicx,bm}
\usepackage{dcolumn}

\begin{document}

\title{Quantum Field Theory of Correlated Bose-Einstein condensates: \\
II. Ward-Takahashi Identities and Correlation Functions}

\author{Takafumi K{\sc ita}}
\affiliation{Department of Physics, Hokkaido University, Sapporo 060-0810, Japan}

\begin{abstract}
We derive Ward-Takahashi identities for correlated Bose-Einstein condensates based on the expressions of the first-order variations $(\delta\Psi,\delta G)$ due to perturbations obtained in the preceding paper [T.\ Kita,\ J.\ Phys.\ Soc.\ Jpn.\ {\bf 90},\ 024001 (2021)] for
the condensate wave function $\Psi$ and Green's function $G$.
They enable us to obtain several exact results on the density and current correlation functions $K_{\nu\nu'}^{}$,
and also express $K_{\nu\nu'}^{}$ in terms of low-energy Green's functions and vertices. 
The latter expressions open up the possibility of constructing theory of superfluid Bose liquids
 in the same way as that for fermions at low temperatures.
The vertices are found to have different limits depending on which of frequency $\omega$ and wavenumber $q$ is set equal to zero first.
\end{abstract}

\maketitle

\section{Introduction}

The Ward identities \cite{Ward50} relates derivatives of Green's function to vertices.
They were extended by Takahashi \cite{Takanashi57} from differential relations 
to finite-difference relations. 

The Ward identities have played a crucial role in providing the Landau theory of Fermi liquids \cite{Landau56,Landau57,Landau58} with a microscopic 
foundation \cite{Pitaevskii59,AGD63,NL62}.
Specifically, they enable us to describe low-temperature properties of a system of homogeneous identical
fermions in terms of low-energy Green's functions and vertices, 
with no reference to the high-energy part of Green's function at all.
They were also used by Leggett \cite{Leggett65,Leggett66} in extending the Landau theory to superfluid phases,
and by Serene and Rainer \cite{SR83} in formulating the quasiclassical theory of superfluid $^3$He
in terms of a few phenomenological Landau parameters. 

In contrast, no derivations of Ward or Ward-Takahashi identities have been known for Bose-Einstein condensates,
except that by Gavoret and Nozi\`eres at zero temperature based on a perturbation expansion \cite{GN64}.
Specifically, their two Ward identities include neither the condensate wave function $\Psi$ nor three-point vertices $\Gamma^{(3)}$ characteristic of Bose-Einstein condensation. 
Moreover, their derivation is based on the fictitious gap method introduced to avoid the infrared divergence in the
simple perturbation expansion, with which the four-point vertices $\Gamma^{(4)}$ are concluded to be identical between the $q$- and $\omega$-limits, i.e., independent of which of the frequency $\omega$ or wavenumber $q$ is set equal to zero first. 
Whether the statements are valid or not should be clarified definitely.
Establishing the identities at finite temperatures, we may also be able to construct a correspondent of the Landau theory of Fermi liquids 
also for Bose-Einstein condensates.

This paper presents a non-perturbative derivation of the Ward-Takahashi identities for Bose-Einstein condensates at finite temperatures.
The basic strategy is outlined as follows.
We apply the local-gauge and truncated-Galilean transformations following Serene and Rainer \cite{SR83}.
On the one hand, we know exactly how the condensate wave function $\Psi$ and Green's function $G$ change under these transformations.
On the other hand, the transformations generate apparent perturbations in the system to yield variations  $(\delta\Psi, \delta G)$,
whose exact expressions with vertices we have already derived in the preceding paper (which is referred to as
I hereafter) \cite{Kita20-1}.
Equating the two results for each perturbation at the first order yields a specific Ward-Takahashi identity.
The resulting Ward-Takahashi identities will be shown to have additional contributions given
in terms of $\Psi$ and
the three-point vertex $\Gamma^{(3)}$ besides those of $G$. 
Moreover, they appropriately reduce in the normal-state limit of $\Psi\rightarrow 0$ 
to the Bose-liquid versions of the four identities well known in the theory of Fermi liquids \cite{AGD63}.

Using them, we will prove several identities obeyed by the density and current correlation functions.
It will also be shown that the correlation functions can be written in terms of
low-energy Green's functions and effective vertices alone. 
This opens up the possibility of constructing low-energy effective theory of correlated Bose-Einstein condensates 
in terms of renormalized Green's functions and vertices, in exactly the same way as for correlated fermions at low temperatures \cite{Pitaevskii59,AGD63,NL62}.

This paper is organized as follows.
Section \ref{sec:2} transforms the expressions of $\delta\Psi$ and $\delta G$ obtained in I
into the energy-momentum space. Section \ref{sec:3} derives Ward-Takahashi identities.
Section \ref{sec:4} introduces the density and current correlation functions, derives several exact results
on them based on the Ward-Takahashi identities, and expresses them in terms of low-energy Green's functions and vertices. Section \ref{sec:5} presents a brief summary.
We set $\hbar=1$ throughout.

\section{Fourier transform of $\delta \Psi$ and $\delta G$\label{sec:2}}

We focus on homogeneous systems throughout. 
We here transform the expressions of first-order variations $(\delta\Psi,\delta G)$ obtained in I
into the energy-momentum space. 
The final results are given in Eq.\ (\ref{dPsidG-sol(q)}) below.

It was shown in I that first-order variations of $(\Psi,G)$ under perturbation
\begin{align}
S_{\rm ext}
\equiv&\, \int dx\int dx'\vec\psi_{}^{\,{\rm T}}(x) \delta\hat{I}(x,x')\vec\psi(x') 
\label{S_ext}
\end{align}
are given exactly by
\begin{subequations}
\label{dPsidG-sol}
\begin{align}
\delta\vec{\Psi}
=&\,\hat{G}\biggl(\delta\hat{I}_{}^{\,({\rm s})}\vec{\Psi}
-\frac{1}{2}\,\underline{\Gamma}_{}^{(3)}\underline{GG}\,\delta \vec{I}_{}^{\,({\rm s})} \biggr)  ,
\label{dPsi-sol}
\\
\delta\vec{G}=&\,\underline{GG}\biggl[\left(\underline{1}-\frac{1}{2}\,\underline{\Gamma}_{}^{(4)}\underline{GG}\right) \delta\vec{I}_{}^{\,({\rm s})} 
+\underline{\Gamma}_{}^{(3){\rm T}}\delta\vec{\Psi}\biggr] ,
\label{dG-sol}
\end{align}
\end{subequations}
with 
\begin{align}
\delta\vec{I}_{}^{\,({\rm s})}(x,x')\equiv \delta\vec{I}(x,x')+\delta\vec{I}_{}^{\,\,{\rm T}}(x',x).
\label{I^(s)}
\end{align}
See Eqs.\ (29), (41), and (42) of I.

Let us expand the $2\times 2$ Nambu Green's function $\hat{G}(x,x')$ in the Fourier series,
\begin{align}
\hat{G}(x,x')=\frac{1}{\beta V}\sum_p \hat{G}(p)\,e^{ip(x-x')} ,
\label{G-exp}
\end{align}
where $V$ is the volume of the system, $x$ and $p$ are defined by $x\!\equiv\! ({\bf r},\tau)$ and 
\begin{align}
p\equiv ({\bf p},\varepsilon_n^{}),
\end{align}
with $\varepsilon_n^{}\!\equiv\! 2\pi n/\beta$
 the boson Matsubara frequency, and $px$ denotes a scalar product
 with the Minkowskii metric given explicitly by
\begin{align}
px\equiv {\bf p}\cdot{\bf r}-\varepsilon_n^{}\tau .
\end{align}
Accordingly, the matrix element of $\underline{GG}$ 
is expressible through a change of variables as
\begin{align}
\langle \xi_1,\xi_1'|\underline{GG}|\xi_2,\xi_2'\rangle \equiv&\,G_{j_1j_2}(x_1,x_2)G_{j_1'j_2'}(x_1',x_2')
\notag \\
=&\, \frac{1}{(\beta V)^2}\sum_{p,q}G_{j_1j_2}^{}(p_+^{})G_{j_1'j_2'}^{}(-p_-^{})
\notag \\
&\,\times e^{ip_+^{}(x_1-x_2)-ip_-^{}(x_1'-x_2')} ,
\label{GG-exp}
\end{align}
where $\xi\equiv (j,x)$ with $j=1,2$, and $q$ and $p_\pm^{}$ are defined by 
\begin{subequations}
\label{p_pm}
\begin{align}
q\equiv &\,({\bf q},\omega_\ell^{}),
\\
p_+^{}\equiv &\,({\bf p}+{\bf q}/2,\varepsilon_n^{}+\omega_\ell^{}), 
\\
p_-^{}\equiv &\,({\bf p}-{\bf q}/2,\varepsilon_n^{}).
\end{align}
\end{subequations}
We also expand the other quantities in Eq.\ (\ref{dPsidG-sol}) as
\begin{subequations}
\label{Gamma-exp}
\begin{align}
\langle \xi |\delta \vec\Psi =&\, \sum_q \delta \Psi_j^{}(q)\,e^{iqx},
\label{dPsi(x)}
\\
\langle \xi,\xi'|\delta \vec{G}=&\,\frac{1}{\beta V}\sum_{pq}\delta G_{jj'} (p;q)\, e^{ip_+^{}x-ip_-^{}x'} ,
\label{dG(x,x')}
\\
\langle \xi,\xi'|\delta \vec{I}^{\,({\rm s})}=&\,\frac{1}{\beta V}\sum_{pq}\delta I_{jj'}^{({\rm s})} (p;q)\, e^{ip_+^{}x-ip_-^{}x'} ,
\label{dI^(x,x')}
\end{align}
\begin{align}
\langle \xi_1,\xi_1'|\underline{\Gamma}^{(4)}|\xi_2,\xi_2'\rangle
=&\, \frac{1}{(\beta V)^3} \sum_{pp'q}\Gamma_{j_1j_1';j_2j_2'}^{(4)}(p,p';q)
\notag \\
&\,\times  e^{ip_+^{}x_1-ip_-^{}x_1'-ip_+'x_2+ip_-'x_2'},
\label{Gamma^(4)-exp}
\end{align}
\begin{align}
\langle \xi_1,\xi_1'|\underline{1}|\xi_2,\xi_2'\rangle
=&\, \frac{\delta_{j_1j_2}^{}\delta_{j_1'j_2'}^{}}{(\beta V)^2} \sum_{pq}e^{ip_+^{}(x_1-x_2)-ip_-^{}(x_1'-x_2')} ,
\end{align}
\begin{align}
\langle \xi_1|\underline{\Gamma}^{(3)}|\xi_2,\xi_2'\rangle
=&\, \frac{1}{(\beta V)^2} \sum_{pq}\Gamma_{j_1;j_2j_2'}^{(3)}(p;q)
\notag \\
&\,\times  e^{iqx_1-ip_+^{}x_2+ip_-^{}x_2'},
\end{align}
\begin{align}
\langle \xi_1,\xi_1'|\underline{\Gamma}^{(3){\rm T}}|\xi_2\rangle
=&\, \frac{1}{(\beta V)^2} \sum_{pq}\Gamma_{j_1j_1';j_2}^{(3){\rm T}}(p;q)
\notag \\
&\,\times  e^{ip_+^{}x_1-ip_-^{}x_1'-iqx_2}.
\end{align}
\end{subequations}

Substituting Eqs.\ (\ref{G-exp}), (\ref{GG-exp}), and (\ref{Gamma-exp}),
we obtain the Fourier transform of Eq.\ (\ref{dPsidG-sol}) as
\begin{subequations}
\label{dPsidG-sol(q)}
\begin{align}
\delta\vec{\Psi}(q)
=&\,\hat{G}(q)\biggl[\delta \hat{I}^{\,({\rm s})}(q_{\rm m}^{}/2\,;q) \vec\Psi 
\notag \\
&\,
-\frac{1}{2\beta V}\sum_p\underline{\Gamma}^{(3)}(p;q) \underline{GG} (p;q) \delta \vec{I}^{\,({\rm s})}(p;q)\biggr]   ,
\label{dPsi-sol(q)}
\\
\delta\vec{G}(p;q) =&\,\underline{GG}(p;q) \Biggl[\delta\vec{I}^{\,({\rm s})}(p;q)
\notag \\
&\,  -\frac{1}{2\beta V}\sum_{p'}\underline{\Gamma}^{(4)}(p,p';q)\underline{GG}(p';q)\delta\vec{I}^{\,({\rm s})}(p';q) 
\notag \\
&\, 
+\,\underline{\Gamma}^{(3){\rm T}}(p;q)\delta\vec{\Psi}(q)\Biggr] ,
\label{dG-sol(q)}
\end{align}
\end{subequations}
where $q_{\rm m}^{}\equiv ({\bf q},0)$ with $_{\rm m}$ denoting {\it momentum}, and $\delta\vec{G}$, $\underline{\Gamma}^{(4)}$, etc., are vectors and matrices in the Nambu (i.e., particle-hole)  space defined by
\begin{subequations}
\label{VecMat(q)}
\begin{align}
\langle j| \vec{\Psi}\equiv&\, \Psi_j,
\\
\langle j| \delta\vec{\Psi}(q)\equiv&\,\delta \Psi_j(q),
\\
\langle j |\hat{G}(q)|j'\rangle\equiv&\, G_{jj'}(q),
\label{<j|hatG|j'>}
\\
\langle jj'| \delta\vec{G}(p;q)\equiv&\,\delta G_{jj'}(p;q),
\label{<jj'|vecG}
\\
\langle jj'| \delta\vec{I}^{\,({\rm s})}(p;q)\equiv&\,\delta I_{jj'}^{({\rm s})}(p;q),
\\
\langle j_1| \underline{\Gamma}^{(3)}(p;q) |j_2j_2'\rangle\equiv&\, \Gamma^{(3)}_{j_1;j_2j_2'}(p;q),
\\
\langle j_1j_1'| \underline{\Gamma}^{(3){\rm T}}(p;q) |j_2\rangle\equiv&\, \Gamma^{(3){\rm T}}_{j_1j_1';j_2}(p;q),
\\
\langle j_1j_1'| \underline{GG}(p;q) |j_2j_2'\rangle\equiv&\, G_{j_1j_2}(p_+^{})G_{j_1'j_2'}(-p_-^{}),
\label{GG(p;q)}
\\
\langle j_1j_1'| \underline{\Gamma}^{(4)}(p,p';q) |j_2j_2'\rangle\equiv&\, \Gamma^{(4)}_{j_1j_1';j_2j_2'}(p,p';q).
\end{align}
\end{subequations}
The symmetries of these vectors and matrices are summarized in Appendix\ref{sec:App1}.

\section{Ward-Takahashi Identities\label{sec:3}}

We derive Ward-Takahashi Identities of Bose-Einstein condensates based on the local-gauge transformation,
truncated-Galilean transformation \cite{SR83}, and a spatially inhomogeneous perturbation \cite{AGD63}.

\subsection{Local gauge transformation\label{subsec3.1}}

First, we perform the local gauge transformation
\begin{align}
\vec\psi^{\,\prime}(x)= e^{i\hat\sigma_3^{}\chi(x)}\,\vec{\psi}(x)
\label{GaugeTrans}
\end{align}
on $\tau\in[0,\beta]$ with periodicity $\chi({\bf r},\tau+\beta)=\chi({\bf r},\tau)$. 
The transformation yields variations $\delta\vec{\Psi}\equiv \vec\Psi'-\vec\Psi$ and $\delta\hat{G}\equiv \hat{G}'-\hat{G}$,
which are given in the first order in $\chi(x)$ by
\begin{subequations}
\label{deltaGPsi}
\begin{align}
\delta \vec{\Psi}(x)=&\, i \hat\sigma_3^{} \chi(x)\vec\Psi,
\\
\delta \hat{G}(x,x')
=&\, i\left[\chi(x)\hat\sigma_3^{} \hat{G}(x,x')+\hat{G}(x,x')\hat\sigma_3^{}\chi(x')\right].
\label{dG-gauge}
\end{align}
\end{subequations}
Let us expand $\chi(x)$ in the Fourier series
\begin{align}
\chi(x)=&\,\sum_q \chi(q)\,e^{iqx} .
\label{chi-exp}
\end{align}
Substituting Eqs.\ (\ref{G-exp}), (\ref{dPsi(x)}), (\ref{dG(x,x')}), and (\ref{chi-exp}) into Eq.\ (\ref{deltaGPsi}), we obtain
\begin{subequations}
\label{deltaGPsi(q)}
\begin{align}
\delta\vec\Psi(q)=&\,  i\chi(q)\hat\sigma_3^{}\vec\Psi ,
\\
\delta \hat{G}(p;q)=&\, i\chi(q)\left[\hat\sigma_3^{}\hat{G}(p_-^{})+\hat{G}(p_+^{})\hat\sigma_3^{}\right] ,
\end{align}
\end{subequations}
where $p_\pm^{}$ are given in Eq.\ (\ref{p_pm}).

On the other hand, transformation (\ref{GaugeTrans}) yields an apparent perturbation in the system. 
Specifically, 
let us substitute $\vec{\psi}(x)\!=\!e^{-i\hat\sigma_3^{}\chi(x)}\vec\psi^{\,\prime}(x)$ into the kinetic part of the action given by
\begin{align}
S_0=&\,\frac{1}{2}\int   d x\,\vec\psi^{\,{\rm T}}(x)\biggr[-i\hat\sigma_2^{}\frac{\partial}{\partial\tau}+\hat\sigma_1^{}\biggl(\frac{\hat{\bf p}^2}{2m} -\mu\biggr)\biggr]\vec\psi(x) ,
\label{S_0}
\end{align}
extract terms first order in $\chi(x)$ and set $\vec\psi^{\,\prime}(x)\rightarrow \vec\psi(x)$ in them
as appropriate in the first order, and expand $\chi(x)$ and $\vec\psi(x)$ as Eq.\ (\ref{chi-exp}) and 
\begin{align}
\vec\psi(x)=\frac{1}{V^{1/2}}\sum_p \vec\psi(p) \,e^{ipx} .
\label{psi-exp}
\end{align}
We thereby find that the perturbation is expressible as
\begin{align}
S_{\rm ext}=&\, \beta \sum_{pq} \vec\psi^{\,{\rm T}}(-p_+^{})\delta \hat{I}(p;q)\vec\psi(p_-^{}) ,
\label{S_ext-2}
\end{align}
with
\begin{align}
\delta \hat{I} (p;q) =\frac{i}{2}\chi(q)\biggl(i\omega_\ell^{} \hat\sigma_1^{}+\frac{{\bf p}\cdot{\bf q}}{m}i\hat\sigma_2^{}\biggr).
\label{I(p;q)}
\end{align}
The corresponding Fourier coefficient of Eq.\ (\ref{I^(s)}) is given by
\begin{align}
\delta \hat{I}^{({\rm s})} (p;q)=\delta \hat{I} (p;q)+\delta \hat{I}^{\,{\rm T}} (-p;q)=2\delta \hat{I} (p;q) .
\label{I^(s)(p;q)}
\end{align}

Let us substitute Eqs.\ (\ref{deltaGPsi(q)}) and (\ref{I(p;q)}) into Eq.\ (\ref{dPsidG-sol(q)}). 
Noting that $\chi(q)$ is arbitrary, we obtain
\begin{subequations}
\label{WT-gauge1}
\begin{align}
\hat\sigma_3^{}\vec\Psi = &\,\hat{G}(q)  \Biggl[\biggl(i\omega_\ell^{} \hat\sigma_1^{}+\frac{{\bf q}^2}{2m}i\hat\sigma_2^{}\biggr)\vec\Psi-\frac{1}{2\beta V}\sum_p
\underline{\Gamma}^{(3)}(p;q)
\notag \\
&\, \times\underline{GG} (p;q)\biggl(i\omega_\ell^{} \vec\sigma_1+\frac{{\bf p}\cdot{\bf q}}{m}i\vec\sigma_2\biggr)\Biggr] ,
\label{WT-gauge11}
\end{align}
\begin{align}
&\,\hat\sigma_3^{}\hat{G}(p_-^{})+\hat{G}(p_+^{})\hat\sigma_3^{}
\notag \\
=&\, \underline{GG}(p;q) \Biggl\{\sum_{p'} \Biggl[\delta_{pp'}\underline{1}-\frac{1}{2\beta V}\underline{\Gamma}^{(4)}(p,p';q)\underline{GG}(p';q)\Biggr]
\notag \\
&\,  \times  \left(i\omega_\ell^{} \vec\sigma_1+\frac{{\bf p}'\cdot{\bf q}}{m}i\vec\sigma_2\right)
+\underline{\Gamma}^{(3){\rm T}}(p;q)\hat\sigma_3^{}\vec{\Psi}\Biggr\},
\label{WT-gauge12}
\end{align}
\end{subequations}
where $\underline{1}$ denotes the $4\times 4$ unit matrix, and  
$\hat{A}=\vec{B}$ in Eq.\ (\ref{WT-gauge12}) signifies the equality $A_{jj'}=B_{jj'}$ between their elements 
defined by $A_{jj'}\equiv\langle j|\hat{A}|j'\rangle$ and $B_{jj'}\equiv \langle jj'|\vec{B}$.
Equation (\ref{WT-gauge1}) is the most general form of the Ward-Takahashi identities concerning the local gauge transformation.
Let us operate $\hat{G}^{-1}(q)$ on Eq.\ (\ref{WT-gauge11}) from the left and write 
\begin{align}
\underline{GG}(p;q)\vec{B}=\hat{G}(p_+^{})\hat{B}\hat{G}(p_-^{}),
\label{GGB}
\end{align}
in Eq.\ (\ref{WT-gauge12}) based on Eqs.\ (\ref{<jj'|vecG}), (\ref{GG(p;q)}), and (\ref{G-symm}).
We can thereby express Eq.\ (\ref{WT-gauge1}) alternatively as
\begin{subequations}
\label{WT-gauge2}
\begin{align}
\hat{G}^{-1}(q)\hat\sigma_3^{}\vec\Psi = &\,  \biggl(i\omega_\ell^{} \hat\sigma_1^{}+\frac{{\bf q}^2}{2m}i\hat\sigma_2^{}\biggr)\vec\Psi-\frac{1}{2\beta V}\sum_p
\underline{\Gamma}^{(3)}(p;q)
\notag \\
&\, \times\underline{GG} (p;q)\biggl(i\omega_\ell^{} \vec\sigma_1+\frac{{\bf p}\cdot{\bf q}}{m}i\vec\sigma_2\biggr) ,
\label{WT-gauge21}
\end{align}
\begin{align}
&\,\hat{G}^{-1}(p_+^{})\hat\sigma_3^{}+\hat\sigma_3^{}\hat{G}^{-1}(p_-^{})
\notag \\
=&\, \sum_{p'} \Biggl[\delta_{pp'}\underline{1}-\frac{1}{2\beta V}\underline{\Gamma}^{(4)}(p,p';q)\underline{GG}(p';q)\Biggr]
\notag \\
&\,  \times  \left(i\omega_\ell^{} \vec\sigma_1+\frac{{\bf p}'\cdot{\bf q}}{m}i\vec\sigma_2\right)
+\underline{\Gamma}^{(3){\rm T}}(p;q)\hat\sigma_3^{}\vec{\Psi}.
\label{WT-gauge22}
\end{align}
\end{subequations}
Note that Eqs.\ (\ref{WT-gauge21}) and (\ref{WT-gauge22}) include the non-interacting limit with $\mu=0$ as trivial equalities, 
as seen by using $\hat{G}_0^{-1}(p)\equiv \varepsilon_n\hat\sigma_2^{}-({\bf p}^2/2m)\hat\sigma_1^{}$.

Setting ${\bf q}={\bf 0}$ in Eq.\ (\ref{WT-gauge2}) yields
\begin{subequations}
\label{WT-gauge3}
\begin{align}
\frac{\hat{G}^{-1}(q_{\rm e}^{})}{i\omega_\ell^{}}\hat\sigma_3^{}\vec\Psi = &\,  \hat\sigma_1^{}\vec\Psi-\frac{1}{2\beta V}\sum_p
\underline{\Gamma}^{(3)}(p;q_{\rm e}^{})\underline{GG} (p;q_{\rm e}^{}) \vec\sigma_1,
\label{WT-gauge31}
\end{align}
\begin{align}
&\,\frac{\hat{G}^{-1}(p_{+{\rm e}}^{})\hat\sigma_3^{}+\hat\sigma_3^{}\hat{G}^{-1}(p)-\underline{\Gamma}^{(3){\rm T}}(p;q_{\rm e}^{})\hat\sigma_3^{}\vec{\Psi}}{i\omega_\ell}
\notag \\
=&\, \hat\sigma_1^{}-\frac{1}{2\beta V}\sum_{p'}\underline{\Gamma}^{(4)}(p,p';q_{\rm e}^{})\underline{GG}(p';q_{\rm e}^{}) \vec\sigma_1,
\label{WT-gauge32}
\end{align}
\end{subequations}
where $q_{\rm e}^{}$ and $p_{+{\rm e}}^{}$ are defined by
\begin{subequations}
\label{q_e}
\begin{align}
q_{\rm e}^{}\equiv &\, (0,\omega_\ell),
\label{q_e1}
\\
p_{+{\rm e}}^{}\equiv&\, ({\bf p},\varepsilon_n+\omega_\ell),
\label{q_e2}
\end{align}
\end{subequations}
with subscript $_{\rm e}$ denoting {\it energy}.
On the other hand, setting $\omega_\ell=0$ in Eq.\ (\ref{WT-gauge2}) yields
\begin{subequations}
\label{WT-gauge4}
\begin{align}
\hat{G}^{-1}(q_{\rm m}^{})\hat\sigma_3^{}\vec\Psi = &\, \frac{{\bf q}^2}{2m}i\hat\sigma_2^{}\vec\Psi-\frac{1}{2\beta V}\sum_p
\underline{\Gamma}^{(3)}(p;q_{\rm m}^{})
\notag \\
&\, \times\underline{GG} (p;q_{\rm m}^{})\frac{{\bf p}\cdot{\bf q}}{m}i\vec\sigma_2 ,
\label{WT-gauge41}
\end{align}
\begin{align}
&\,\hat{G}^{-1}(p_{+{\rm m}}^{})\hat\sigma_3^{}+\hat\sigma_3^{}\hat{G}^{-1}(p_{-{\rm m}}^{})
\notag \\
=&\, \frac{{\bf p}\cdot{\bf q}}{m}i\hat\sigma_2^{} -\frac{1}{2\beta V}\sum_{p'}\underline{\Gamma}^{(4)}(p,p';q_{\rm m}^{})\underline{GG}(p';q_{\rm m}^{})
\frac{{\bf p}'\cdot{\bf q}}{m}i\vec\sigma_2
\notag \\
&\, 
+\underline{\Gamma}^{(3){\rm T}}(p;q_{\rm m}^{})\hat\sigma_3^{}\vec{\Psi},
\label{WT-gauge42}
\end{align}
\end{subequations}
where $q_{\rm m}^{}$ and $p_{+{\rm m}}^{}$ are defined by
\begin{subequations}
\label{q_m}
\begin{align}
q_{\rm m}^{}\equiv &\, ({\bf q},0),
\label{q_m1}
\\
p_{+{\rm m}}^{}\equiv &\,({\bf p}\pm{\bf q}/2,\varepsilon_n),
\label{q_m2}
\end{align}
\end{subequations}
with subscript $_{\rm m}$ denoting {\it momentum}.
Taking the limit $\omega_\ell^{}\rightarrow 0$ in Eq.\ (\ref{WT-gauge3}) and differentiating Eq.\ (\ref{WT-gauge4}) 
with respect to ${\bf q}$, we obtain the Ward identities from the local gauge transformation, which
for $\vec\Psi\!\rightarrow\! \vec 0$ reduce to the expressions
essentially identical with those 
well-known in the theory of normal Fermi liquids, i.e., Eqs.\ (19.1) and (19.2) of Abrikosov, Gor'kov, and Dzyaloshinski \cite{AGD63}.

\subsection{Truncated Galilean transformation}

Next, we consider the truncated Galilean transformation \cite{SR83}
\begin{align}
\vec\psi^{\,\prime}(x)=\vec\psi(x+X(\tau)),
\label{Galilean}
\end{align}
where $X(\tau)\!\equiv\! ({\bf R}(\tau),0)$ with ${\bf R}(\tau+\beta)\!=\!{\bf R}(\tau)$.
The condensate wave function 
remains invariant through the transformation as it does not depend on $x$ for homogeneous systems, i.e.,
\begin{align}
\delta\vec\Psi=0 .
\label{dPsi-G}
\end{align}
In contrast, variation $\delta\hat{G}\!\equiv\! \hat{G}'-\hat{G}$ in Green's function
is finite, which in the first order is given by
\begin{align}
\delta\hat{G}(x,x')= \left[{\bf R}(\tau)\cdot{\nabla}+{\bf R}(\tau')\cdot{\bf \nabla}'\right]\hat{G}(x,x').
\label{deltaG-G}
\end{align}
Let us expand ${\bf R}(\tau)$ in the Fourier series
\begin{align}
{\bf R}(\tau)=&\,\sum_\ell {\bf R}(q_{\rm e}^{})\,e^{iq_{\rm e}^{}x} ,
\label{R-exp}
\end{align}
with $q_{\rm e}^{}$ defined by Eq.\ (\ref{q_e1}).
Substituting Eqs.\ (\ref{G-exp}), (\ref{dG(x,x')}), and (\ref{R-exp}) into Eq.\ (\ref{deltaG-G}), we obtain
\begin{align}
\delta \hat{G}(p;q)=&\, i{\bf R}(\omega_\ell)\cdot{\bf p} \left[\hat{G}(p)-\hat{G}(p_{+{\rm e}}^{})\right] ,
\label{deltaG(q)-G}
\end{align}
where $p_{+{\rm e}}^{}$ is defined by Eq.\ (\ref{q_e2}).

On the other hand, transformation (\ref{Galilean}) yields an apparent perturbation in the system. 
Specifically, 
let us substitute $\vec\psi(x)=\vec\psi^{\,\prime}(x-X(\tau))$ into Eq.\ (\ref{S_0}),
extract terms first-order in ${\bf R}(\tau)$ and set $\vec\psi^{\,\prime}(x)\rightarrow \vec\psi(x)$ in them
as appropriate in the first order, and expand ${\bf R}(\tau)$ and $\vec\psi(x)$ as Eqs.\ (\ref{R-exp}) and 
(\ref{psi-exp}), respectively.
We thereby find that the perturbation is expressible as Eq.\ (\ref{S_ext-2}) with
\begin{align}
\delta \hat{I} (p;q) =\frac{\delta_{{\bf q}{\bf 0}}}{2}\omega_\ell {\bf R}(\omega_\ell)\cdot {\bf p}\, i \hat\sigma_2^{},
\label{I(p;q)-2}
\end{align}
which satisfies Eq.\ (\ref{I^(s)(p;q)}).

Let us substitute Eqs.\ (\ref{dPsi-G}), (\ref{deltaG(q)-G}), and (\ref{I(p;q)-2}) into Eq.\ (\ref{dPsidG-sol(q)}). 
Noting that ${\bf R}(\omega_\ell)$ is arbitrary, we obtain
\begin{subequations}
\label{WT-Galilean}
\begin{align}
\hat{G}(q_{\rm e})\frac{1}{2\beta V}\sum_p\underline{\Gamma}^{(3)}(p;q_{\rm e}^{}) \underline{GG} (p;q_{\rm e}^{}) {\bf p}\, i \vec\sigma_2=0,
\label{WT-Galilean1}
\end{align}
\begin{align}
&\,{\bf p} \frac{\hat{G}(p_{+{\rm e}}^{})-\hat{G}(p)}{i\omega_\ell}
\notag \\
=&\, \underline{GG}(p;q_{\rm e}^{}) \Biggl[ {\bf p}\, i \vec\sigma_2  -\frac{1}{2\beta V}\sum_{p'} \underline{\Gamma}^{(4)}(p,p';q_{\rm e}^{})
\underline{GG}(p';q_{\rm e}^{})
\notag \\
&\,\times  {\bf p}'\, i \vec\sigma_2\Biggr].
\label{WT-Galilean2}
\end{align}
\end{subequations}
These identities are expressible alternatively by operating $\hat{G}^{-1}(q_{\rm e})$ on Eq.\ (\ref{WT-Galilean1}) from the left and using Eq.\ (\ref{GGB}) 
in Eq.\ (\ref{WT-Galilean2}) as
\begin{subequations}
\label{WT-Galilean-a}
\begin{align}
\frac{1}{2\beta V}\sum_p\underline{\Gamma}^{(3)}(p;q_{\rm e}^{}) \underline{GG} (p;q_{\rm e}^{}) {\bf p}\, i \vec\sigma_2=0,
\label{WT-Galilean1a}
\end{align}
\begin{align}
&\,\frac{{\bf p}}{m} \frac{\hat{G}^{-1}(p)-\hat{G}^{-1}(p_{+{\rm e}}^{})}{i\omega_\ell}
\notag \\
=&\, \frac{{\bf p}}{m}\, i \hat\sigma_2^{}  -\frac{1}{2\beta V}\sum_{p'} \underline{\Gamma}^{(4)}(p,p';q_{\rm e}^{})
\underline{GG}(p';q_{\rm e}^{})
\frac{{\bf p}'}{m} i \vec\sigma_2 .
\label{WT-Galilean2a}
\end{align}
\end{subequations}
Equation (\ref{WT-Galilean2a}) in the limit of $\omega_\ell^{}\rightarrow 0$ is essentially identical with
Eq.\ (19.3) in Ref.\ \onlinecite{AGD63} for normal Fermi liquids.

\subsection{Spatially inhomogeneous perturbation}

Finally, we apply a static inhomogeneous potential to the system $U_{\rm ext}({\bf r})$. 
Let us expand 
\begin{align}
U_{\rm ext}({\bf r})= \sum_q U_{\rm ext}({\bf q}) \,e^{iq_{\rm m}^{}x} ,
\end{align}
with $q_{\rm m}^{}$ defined by Eq.\ (\ref{q_m1}).
The perturbation is expressible as Eq.\ (\ref{S_ext-2}) with
\begin{align}
\delta \hat{I} (p;q) =\frac{1}{2} \delta_{\omega_\ell^{} 0} U_{\rm ext}({\bf q})\hat\sigma_1^{},
\label{I(p;q)-3}
\end{align}
which also satisfies Eq.\ (\ref{I^(s)(p;q)}).

In the limit of ${\bf q}\rightarrow{\bf 0}$, the perturbation can be absorbed into the shift of the chemical potential
given by
$\delta \mu = -U_{\rm ext}({\bf 0})$.
Let us substitute Eq.\ (\ref{I(p;q)-3}) into Eq.\ (\ref{dPsidG-sol(q)}), take the limit ${\bf q}\rightarrow{\bf 0}$, 
and replace $U_{\rm ext}({\bf 0})$ by $-\delta \mu$.
We thereby obtain
\begin{subequations}
\label{dPsiG/dmu1}
\begin{align}
\frac{\partial\vec\Psi}{\partial\mu}=&\,-\lim_{q_{\rm m}^{}\rightarrow 0}\hat{G}(q_{\rm m}^{})\biggl[\hat\sigma_1^{}\vec\Psi 
\notag \\
&\,- \frac{1}{2\beta V}
\sum_p\underline{\Gamma}^{(3)}(p;q_{\rm m}^{})  \underline{GG} (p;q_{\rm m}^{}) \vec\sigma_1\biggr],
\label{dPsiG/dmu11}
\end{align}
\begin{align}
\frac{\partial \vec{G}(p)}{\partial\mu}= &\,-\!
\underline{GG}^q(p) \Biggl[\vec\sigma_1
 -\frac{1}{2\beta V}\sum_{p'}\underline{\Gamma}^{(4q)}(p,p')\underline{GG}^q(p')\vec\sigma_1
\notag \\
&\, 
-\underline{\Gamma}^{(3q){\rm T}}(p)\frac{\partial\vec{\Psi}}{\partial\mu}\Biggr] ,
\label{dPsiG/dmu12}
\end{align}
\end{subequations}
where superscript $^q$ denotes the $q$-limit of setting $\omega_\ell=0$ first and taking the limit ${\bf q}\rightarrow{\bf 0}$
subsequently \cite{Landau58,AGD63,SR83}, e.g., 
\begin{align}
\underline{\Gamma}^{(4q)}(p,p')\equiv \lim_{q_{\rm m}^{}\rightarrow 0}
\underline{\Gamma}^{(4)}(p,p';q_{\rm m}^{}).
\label{q-limit}
\end{align}
Equation (\ref{dPsiG/dmu12}) is expressible alternatively by using Eq.\ (\ref{GGB}) and $\delta G^{-1}=-G^{-1}\delta G \,G^{-1}$ as
\begin{align}
\frac{\partial \vec{G}^{-1}(p)}{\partial\mu}= &\,
\vec\sigma_1
 -\frac{1}{2\beta V}\sum_{p'}\underline{\Gamma}^{(4q)}(p,p')\underline{GG}^q(p')\vec\sigma_1
\notag \\
&\, 
-\underline{\Gamma}^{(3q){\rm T}}(p)\frac{\partial\vec{\Psi}}{\partial\mu} .
\label{dPsiG^-1/dmu12}
\end{align}
Equation (\ref{dPsiG^-1/dmu12}) is the superfluid-Bose-liquid version of
Eq.\ (19.4) in Ref.\ \onlinecite{AGD63} for normal Fermi liquids.

Derivative $\partial\vec\Psi/\partial\mu$ should be finite, which implies that the quantity in the square brackets of Eq.\ (\ref{dPsiG/dmu11}) should vanish in the $q$-limit, i.e.,
\begin{align}
\hat\sigma_1^{}\vec\Psi 
- \frac{1}{2\beta V}
\sum_p\underline{\Gamma}^{(3q)}(p)  \underline{GG}^q (p) \vec\sigma_1=0 .
\label{identity-q-limit}
\end{align}
It is shown in Appendix\ref{sec:App2} that Eq.\ (\ref{identity-q-limit}) can also be derived directly from Eq.\ (\ref{dPsiG^-1/dmu12}).

\section{Density and Current Correlation Functions\label{sec:4}}

In this section, we derive formally exact expressions of the density and current correlation functions, i.e., 
Eq.\ (\ref{K_nn'(q)-2}) below, 
based on the linear-response treatment under external perturbations.
We then clarify their properties and also derive low-energy expressions using the Ward-Takahashi identities.

\subsection{Derivation}

Consider the perturbation
\begin{align}
S_{\rm ext}=\beta \sum_{q(\neq 0)} \sum_{\nu=1}^4 j_\nu^{} (-q)A_\nu^{}(q) ,
\label{S'}
\end{align}
where $j_\nu^{} (-q)$ is defined by
\begin{align}
j_\nu^{} (-q)=\frac{1}{2}\sum_p\vec\psi^{\,{\rm T}}(-p_+^{})\hat\lambda_\nu({\bf p})\vec\psi(p_-^{}), 
\label{j_nu}
\end{align}
in terms of vertices
\begin{align}
\hat\lambda_\nu^{}({\bf p})\equiv 
\left\{\begin{array}{ll}
\vspace{2mm}
\displaystyle \frac{p_\nu^{}}{m}i\hat\sigma_2^{}  & : \nu=1,2,3
\\
\hat\sigma_1^{} & : \nu=4 
\end{array}\right. 
\label{lambda_nu}
\end{align}
satisfying $\hat\lambda_\nu^{}({\bf p})=\hat\lambda_\nu^{\rm T}(-{\bf p})$.
Equation (\ref{S'}) represents a perturbation linear in $A_\nu^{}(q)$ obtained from Eq.\ (\ref{S_0}) by (i) adding the scalar potential $A_4(x)$
in the square brackets,
(ii) replacing $\hat{\bf p}\rightarrow \hat{\bf p}+{\bf A}$, and (iii) expanding $A_\nu^{}(x)=\sum_q A_\nu(q)e^{iqx}$. 

The thermodynamic average of $j_\nu^{}(q)$, which vanishes without the perturbation, can be written 
up to the linear order in $A_\nu^{}$ as
\begin{align}
\langle j_\nu^{}(q)\rangle_A^{} =&\, \frac{1}{2}\sum_p\langle \vec\psi^{\,{\rm T}}(-p_-^{})\hat\lambda_\nu({\bf p})\vec\psi(p_+^{})\rangle_A^{}
\notag \\
=&\, \frac{V}{2}\left[\vec\Psi^{\,\rm T}\hat\lambda_\nu\!\biggl(\frac{{\bf q}}{2}\biggr)\delta \vec\Psi(q)+\delta\vec\Psi^{\,\rm T}(q)\hat\lambda_\nu\!\biggl(\!-\frac{{\bf q}}{2}\biggr)\vec\Psi\right]
\notag \\
&\,-\frac{1}{2\beta}{\sum_{p}}'\vec\lambda_\nu^{\,{\rm T}}(-{\bf p})\delta \vec{G}(p;q) ,
\label{j_nu(q)}
\end{align}
where subscript $_A$ denotes the thermodynamic average under the perturbation, 
and the primed sum over $p$ signifies excluding the states $p_\pm^{}\!=\! 0$. 
In deriving the second expression, we have (i) used that
\begin{align*}
\langle \vec\psi(p)\rangle \approx V^{1/2}\bigl[\delta_{p0}\vec\Psi+\delta_{pq}\,\delta\vec\Psi(q)\bigr]
\end{align*}
holds up to the first order, (ii) performed the transformation for $p_\pm^{}\!\neq\! 0$:
\begin{align*}
&\,\langle\vec\psi^{\,{\rm T}}(-p_-^{})\hat\lambda_\nu({\bf p})\vec\psi(p_+^{})\rangle_A^{}= {\rm Tr}\,
\hat\lambda_\nu({\bf p})\langle \vec\psi(p_+^{})\vec\psi^{\,{\rm T}}(-p_-^{})\rangle_A^{} 
\notag \\
&\,= -\beta^{-1}{\rm Tr}\,
\hat\lambda_\nu({\bf p}) \delta \hat{G}(p;q)= -\beta^{-1}{\rm Tr}\,
\hat\lambda_\nu^{\rm T}(-{\bf p}) \delta \hat{G}(p;q),
\end{align*}
and (iii) expressed the trace of the product of $2\times 2$ matrices $\hat{C}^{\rm T}$ and $\hat{D}$ as
\begin{align}
{\rm Tr}\, \hat{C}^{\rm T}\hat{D} = \sum_{jj'}C_{j'j}D_{j'j} =\vec{C}^{\,{\rm T}}\vec{D} ,
\label{TrAB}
\end{align}
where $\vec{C}^{\,{\rm T}}$ and $\vec{D}$ are defined by
\begin{align}
\vec{C}^{\,{\rm T}}|jj'\rangle =C_{jj'} ,\hspace{5mm} \langle jj'|\vec{D}=D_{jj'} .
\label{vecA^TBdef}
\end{align}
Note that the two terms in the square brackets of Eq.\ (\ref{j_nu(q)}) yield the same contribution, as can be checked directly by using Eq.\ (\ref{lambda_nu}).

Our correlation functions are defined in terms of Eq.\ (\ref{j_nu(q)}) by
\begin{align}
K_{\nu\nu'}^{}(q)\equiv \left.\frac{\langle j_\nu^{}(q)\rangle_A^{}}{\delta A_{\nu'}^{}(q)}\right|_{A=0} = -\beta \langle j_\nu^{}(q) j_{\nu'}^{}(-q)\rangle ,
\label{K_nn'}
\end{align}
where the second expression results directly from the definition of the thermodynamic average in the presence of the perturbation $H_A^{}\!\equiv\! S_{\rm ext}/\beta$.
Comparing Eq.\ (\ref{S'}) with Eq.\ (\ref{S_ext-2}), 
we see the correspondence $\hat{I}(p;q)=\frac{1}{2}\sum_\nu\hat\lambda_\nu({\bf p})A_\nu^{}(q)$ so that Eq.\ (\ref{I^(s)(p;q)}) is now given by
\begin{align}
\hat{I}^{({\rm s})}(p;q)=\sum_\nu\hat\lambda_\nu({\bf p})A_\nu^{}(q).
\label{I^(s)-A}
\end{align}
Let us substitute Eq.\ (\ref{dPsidG-sol(q)}) with Eq.\ (\ref{I^(s)-A}) into Eq.\ (\ref{j_nu(q)}) and perform the differentiation of Eq.\ (\ref{K_nn'}) to obtain $K_{\nu\nu'}(q)$.
The result is expressible symbolically as
\begin{align}
K_{\nu\nu'}(q)=&\, - \frac{1}{2\beta} \,\vec\lambda_\nu^{\,{\rm T}}\underline{GG}\left(\underline{1}-\frac{1}{2\beta V}\underline{\Gamma}^{(4)}\underline{GG}\right)\vec\lambda_{\nu'}
\notag \\
&\,
+ V\left[ \vec\Psi^{\rm T}\hat\lambda_\nu({\bf q}/2)-\frac{1}{2\beta V} \, \vec\lambda_\nu^{\,{\rm T}}\underline{GG}\,\underline{\Gamma}^{(3){\rm T}}\right]\hat{G}(q)
\notag \\
&\,\times  \left[\hat\lambda_{\nu'}({\bf q}/2)\vec\Psi
-\frac{1}{2\beta V}\underline{\Gamma}^{(3)}\underline{GG}\,\vec\lambda_{\nu'}\right] .
\label{K_nn'-3}
\end{align}
To express it more concisely and specifically, we introduce renormalized vertices $\hat\Lambda_\nu^{(2)}(q) $ and $\vec\Lambda_{\nu}^{(4)}(p;q)$ by
\begin{subequations}
\label{Lambda^(4,2)}
\begin{align}
\hat\Lambda_{\nu}^{(2)}(q) \vec\Psi\equiv&\,\hat\lambda_\nu\biggl(\frac{\bf q}{2}\biggr)\vec\Psi
-\!\frac{1}{2\beta V}\!\sum_{p}\underline{\Gamma}^{(3)}(p;q)
\underline{GG}(p;q)\vec\lambda_{\nu}({\bf p}),
\label{Lambda^(2)}
\end{align}
\begin{align}
\vec\Lambda_{\nu}^{(4)}(p;q)\equiv &\,\vec\lambda_{\nu}({\bf p})
\!-\!\frac{1}{2\beta V}\!\sum_{p'}\underline{\Gamma}^{(4)}(p,p';q)\underline{GG}(p';q)\vec\lambda_{\nu}({\bf p}').
\label{Lambda^(4)}
\end{align}
\end{subequations}
Equation (\ref{K_nn'-3}) can be written in terms of them as
\begin{subequations}
\label{K_nn'(q)-2}
\begin{align}
K_{\nu\nu'}(q)
=&\,- \frac{1}{2\beta}\sum_{p}\vec\lambda_\nu^{\,{\rm T}}(-{\bf p})\underline{GG}(p;q)\vec\Lambda_{\nu'}^{(4)}(p;q) 
\notag \\
&\, + V\bigl[\hat\Lambda_\nu^{(2)}(-q)\vec\Psi\bigr]^{\rm T} \hat{G}(q)\,\hat\Lambda_{\nu'}^{(2)}(q) \vec\Psi,
\label{K_nn'(q)-21}
\end{align}
or alternatively,
\begin{align}
K_{\nu\nu'}(q)
=&\,- \frac{1}{2\beta}\sum_{p}\vec\Lambda_{\nu}^{(4){\rm T}}(-p;-q) \underline{GG}(p;q)\vec\lambda_{\nu'}({\bf p})
\notag \\
&\, + V\bigl[\hat\Lambda_\nu^{(2)}(-q)\vec\Psi\bigr]^{\rm T} \hat{G}(q)\,\hat\Lambda_{\nu'}^{(2)}(q) \vec\Psi ,
\label{K_nn'(q)-22}
\end{align}
\end{subequations}
which is the key result here.
Here we have used the identity on Eq.\ (\ref{Lambda^(4,2)}),
\begin{subequations}
\begin{align}
&\, \vec\Psi^{\rm T}\hat\lambda_\nu({\bf q}/2)-\frac{1}{2\beta V}\sum_p
\vec\lambda_\nu^{\,{\rm T}}(-{\bf p})\underline{GG}(p;q)\underline{\Gamma}^{(3){\rm T}}(p;q)
\notag \\
=&\, \bigl[\hat\Lambda_\nu^{(2)}(-q) \vec\Psi\bigr]^{\rm T},
\label{Lambda^(2)T}
\end{align}
\begin{align}
&\,\sum_{p'}\vec\lambda_\nu^{\,{\rm T}}(-{\bf p}')\left[\underline{1}\delta_{pp'}-\frac{1}{2\beta V}\underline{GG}(p';q)\underline{\Gamma}^{(4)}(p',p;q)\right]|jj'\rangle
\notag \\
=&\, \langle jj'|\vec\Lambda_\nu^{(4)}(-p;-q)
\equiv \vec\Lambda_\nu^{(4){\rm T}}(-p;-q)|jj'\rangle,
\label{Lambda^(4)T}
\end{align}
\end{subequations}
the latter of which is given in the notation of Eq.\ (\ref{vecA^TBdef}); they can be shown to hold by using the symmetries given in Appendix\ref{sec:App1}. 

Two comments are in order concerning the correlation functions.
First, Eq.\ (\ref{K_nn'(q)-2}) has the structure obtained by Gavoret and Nozi\`eres at $T=0$,
where its first and second terms on the right-hand side are called the {\it regular} and {\it singular}
 parts, respectively.
Thus, the structure persists also at finite temperatures.
Second, Eq.\ (\ref{K_nn'(q)-2}) can be regarded as the second derivative
$$K_{\nu\nu'}(q)=\frac{1}{2}\frac{\delta^2 \Delta\Omega}{\delta A_\nu^{}(-q)\delta A_{\nu'}^{}(q)}$$
of the variation $\Delta\Omega$ in the grand potential due to the perturbation.
Noting Eqs.\ (\ref{dPsi-sol(q)}), (\ref{I^(s)-A}), and (\ref{K_nn'-3}), we can identify $\Delta\Omega$ as
\begin{align}
\Delta\Omega=&\, - \frac{1}{2\beta} \sum_{\nu\nu'}\sum_{pq}
A_\nu^{}(-q)\vec\lambda_\nu^{\,{\rm T}}(-{\bf p})\underline{GG}(p;q)\vec\Lambda_{\nu'}^{(4)}(p;q)
\notag \\
&\, \times  A_{\nu'}^{}(q)+\sum_q \delta\vec\Psi^{\,{\rm T}}(-q) \hat{G}^{-1}(q)\delta\vec\Psi(q) .
\label{DeltaOmega}
\end{align}
Thus, $\Delta\Omega$ cannot be expressible in terms of the change $\delta\vec\Psi(q)$ of the condensate wave function alone, 
contrary to the considerations by Baym \cite{Baym69} and Holtzmann and Baym \cite{HB07},
but there is additional one from the excited states given by the first term on the right-hand side,
which Gavoret and Nozi\'eres call the {\it regular part} \cite{GN64}.
Moreover, it follows from the isotropy of the system that $\delta\Psi(q)$ for $q=q_{\rm m}^{}\equiv ({\bf q},0)$ is proportional to ${\bf q}\cdot{\bf A}(q_{\rm m}^{})$
so that the second term can only contribute to the longitudinal part of the current correlation functions.
Thus, any microscopic study on the superfluid density, which is relevant to the transverse response, 
should be based on the first term with $\underline{GG}$ in Eq.\ (\ref{DeltaOmega}) instead of the second term with $\delta\vec\Psi$.
In this connection, the ``proofs'' of the Josephson sum rule \cite{Baym69,HB07} and the inequality $-G({\bf p},0)\geq mV\Psi^2/Np^2$ \cite{Baym69}, which rely on the 
second term of Eq.\ (\ref{DeltaOmega}) alone, cannot be justified \cite{comment}.
Neither can the claim by Watabe \cite{Watabe20} based on these sum rule and inequality against 
the emergence of the anomalous exponent in $G$ predicted by our recent renormalization group study \cite{Kita19-1,Kita19-2}.

\subsection{Properties of renormalized vertices}

It follows from the Ward-Takahashi identities that the renormalized vertices in Eq.\ (\ref{Lambda^(4,2)}) satisfy several identities.
First, we replace $q$ by $q_{\rm e}^{}\equiv ({\bf 0},\omega_\ell)$ in Eqs.\ (\ref{Lambda^(2)}) and (\ref{Lambda^(4)}),
set $\nu=4$ and substitute $\hat\lambda_4=\hat\sigma_1^{}$ from Eq.\ (\ref{lambda_nu}),
and compare the resulting expressions with Eqs.\ (\ref{WT-gauge31}) and (\ref{WT-gauge32}), respectively. 
We thereby obtain
\begin{subequations}
\label{Lambda^(4,2)_4(q_e)}
\begin{align}
\hat\Lambda_4^{(2)}(q_{\rm e}^{}) \vec\Psi=&\,\frac{\hat{G}^{-1}(q_{\rm e}^{})}{i\omega_\ell^{}}\hat\sigma_3^{}\vec\Psi ,
\label{Lambda^(2)_4(q_e)}
\end{align}
\begin{align}
\vec\Lambda_4^{(4)}(p;q_{\rm e}^{})= &\,\frac{\hat{G}^{-1}(p_{+{\rm e}}^{})\hat\sigma_3^{}+\hat\sigma_3^{}\hat{G}^{-1}(p)-\underline{\Gamma}^{(3){\rm T}}(p;q_{\rm e}^{})\hat\sigma_3^{}\vec{\Psi}}{i\omega_\ell} .
\label{Lambda^(4)_4(q_e)}
\end{align}
\end{subequations}
It should be noted that the numerator on the right-hand of Eq.\ (\ref{Lambda^(4)_4(q_e)}) certainly vanishes 
in the limit of $\omega_\ell\rightarrow 0$ according to Eq.\ (19) of I.

Second, we replace $q$ by $q_{\rm m}^{}\equiv ({\bf q},0)$ in Eqs.\ (\ref{Lambda^(2)}) and (\ref{Lambda^(4)}),
multiply it by $q_\nu^{}$, take the sum over $\nu=1,2,3$ and substitute $\hat\lambda_\nu=(p_\nu^{}/m)i\hat\sigma_2^{}$ from Eq.\ (\ref{lambda_nu}), and compare the resulting expressions with Eqs.\ (\ref{WT-gauge41}) and (\ref{WT-gauge42}), 
respectively. 
We thereby obtain
\begin{subequations}
\label{Lambda^(4,2)(q_m)}
\begin{align}
\sum_{\nu=1}^3 q_\nu \hat\Lambda_{\nu}^{(2)}(q_{\rm m}^{}) \vec\Psi=&\,\hat{G}^{-1}(q_{\rm m}^{})\hat\sigma_3^{}\vec\Psi ,
\label{Lambda^(2)(q_m)}
\\
\sum_{\nu=1}^3 q_\nu \vec\Lambda_{\nu}^{(4)}(p;q_{\rm m}^{})= &\,\hat{G}^{-1}(p_{+{\rm m}}^{})\hat\sigma_3^{}+\hat\sigma_3^{}\hat{G}^{-1}(p_{-{\rm m}}^{})
\notag \\
&\,-\underline{\Gamma}^{(3){\rm T}}(p;q_{\rm m}^{})\hat\sigma_3^{}\vec{\Psi}.
\label{Lambda^(4)(q_m)}
\end{align}
Alternatively, we differentiate Eq.\ (\ref{WT-gauge42}) with respect to $q_\nu^{}$ for $\nu=1,2,3$, take the limit of ${\bf q}\rightarrow{\bf 0}$,
and compare the resulting expression with Eq.\ (\ref{Lambda^(4)}) in the $q$-limit
defined generally by Eq.\ (\ref{q-limit}).
We thereby obtain
\begin{align}
\vec\Lambda_{\nu}^{(4q)}(p)= &\,\frac{\partial}{\partial q_\nu^{}}\Bigl[
\hat{G}^{-1}(p_{+{\rm m}}^{})\hat\sigma_3^{}+\hat\sigma_3^{}\hat{G}^{-1}(p_{-{\rm m}}^{})
\notag \\
&\,
-\underline{\Gamma}^{(3){\rm T}}(p;q_{\rm m}^{})\hat\sigma_3^{}\vec{\Psi}\Bigr]_{{\bf q}={\bf 0}},
\label{Lambda^(4)(q_m)-2}
\end{align}
\end{subequations}
for $\nu=1,2,3$.

Third, we replace $q$ by $q_{\rm e}^{}$ in Eqs.\ (\ref{Lambda^(2)}) and (\ref{Lambda^(4)}), set $\nu=1,2,3$ and
substitute $\hat\lambda_\nu(p)=(p_\nu^{}/m)i\hat\sigma_2^{}$ from Eq.\ (\ref{lambda_nu}),
and compare the resulting expressions with Eqs.\ (\ref{WT-Galilean1}) and (\ref{WT-Galilean2a}). We thereby obtain
\begin{subequations}
\label{Lambda^(4,2)_j(q_e)}
\begin{align}
\hat\Lambda_{\nu}^{(2)}(q_{\rm e}^{}) \vec\Psi=&\,\vec{0},
\label{Lambda^(2)_j(q_e)}
\end{align}
\begin{align}
\vec\Lambda_{\nu}^{(4)}(p;q_{\rm e}^{})= &\,\frac{p_\nu^{}}{m}\frac{\vec{G}^{-1}(p)-\vec{G}^{-1}(p_{+{\rm e}}^{})}{i\omega_\ell} ,
\label{Lambda^(4)_j(q_e)}
\end{align}
\end{subequations}
for $\nu=1,2,3$.

Fourth, we set $\nu=4$ in Eqs.\ (\ref{Lambda^(2)}) and (\ref{Lambda^(4)}) and take their $q$-limits. 
Comparing the resulting expressions with Eqs.\ (\ref{identity-q-limit}) and (\ref{dPsiG^-1/dmu12}), respectively, we obtain
\begin{subequations}
\label{Lambda^(4q,2q)_4}
\begin{align}
\hat\Lambda_4^{(2q)}\vec\Psi=\vec{0},
\label{Lambda^(2q)_4}
\end{align}
\begin{align}
\vec\Lambda_4^{(4q)}(p)=\frac{\partial \vec{G}^{-1}(p)}{\partial\mu}+\underline{\Gamma}^{(3q){\rm T}}(p)\frac{\partial\vec{\Psi}}{\partial\mu} .
\label{Lambda^(4q)_4}
\end{align}
\end{subequations}

\subsection{Properties of $K_{\nu\nu'}^{}(q)$} 

Let us enumerate properties of Eq.\ (\ref{K_nn'(q)-2}),
which are given by Eqs.\ (\ref{K_nn'-symm})-(\ref{CSR}) below.

First, Eq.\ (\ref{K_nn'}) satisfies
\begin{align}
K_{\nu\nu'}^{}(q)=^{}K_{\nu'\nu}(-q) ,
\label{K_nn'-symm}
\end{align}
as seen from Eq.\ (\ref{K_nn'}).
One can also confirm based on Eq.\ (\ref{Lambda^(2)T})
and the symmetries given in Appendix\ref{sec:App1}
that Eq.\ (\ref{K_nn'(q)-2}) satisfies Eq.\ (\ref{K_nn'-symm}).

Second, $K_{\nu\nu'}(q_{\rm e}^{})$ with $q_{\rm e}^{}=({\bf 0},\omega_\ell)$ vanishes identically, i.e.,
\begin{align}
K_{\nu\nu'}(q_{\rm e}^{})=0.
\label{K_nn'(q_e)=0}
\end{align}
This is proved in the three steps of (i)-(iii) below:
(i) We set $\nu'=1,2,3$  and $q= q_{\rm e}^{}$ in Eq.\ (\ref{K_nn'(q)-21}), 
substitute Eq.\ (\ref{Lambda^(4,2)_j(q_e)}),
and transform the resulting expression as follows:
\begin{align*}
&\,K_{\nu {\nu'}}^{}(q_{\rm e}^{})
\notag \\
=&\, - \frac{1}{2\beta}\sum_{p}\vec\lambda_\nu^{\,{\rm T}}(-{\bf p})\underline{GG}(p;q_{\rm e}^{})
\frac{\vec{G}^{-1}(p)-\vec{G}^{-1}(p_{+{\rm e}}^{})}{i\omega_\ell} \,\frac{p_{\nu'}^{}}{m}
\notag \\
=&\, - \frac{1}{2\beta}\sum_{p}{\rm Tr}\, \hat\lambda_\nu^{\,{\rm T}}({\bf p})
\frac{\hat{G}(p_{+{\rm e}}^{})-\hat{G}(p)}{i\omega_\ell} \,\frac{p_{\nu'}^{}}{m} ,
\end{align*}
where we have used Eq.\ (\ref{GGB}) and $\vec\lambda_\nu^{\,{\rm T}}(-{\bf p})\vec{B}={\rm Tr}\,
\hat\lambda_\nu^{}({\bf p})\hat{B}$.
The last expression can be shown to vanish for each of $\nu=1,2,3$ and $\nu=4$ by making a change of variables $p_{+{\rm e}}^{}\rightarrow p$ in the summation over $\hat{G}(p_{+{\rm e}}^{})$, which does not affect the vertex of Eq.\ (\ref{lambda_nu}).
Thus, we arrive at Eq.\ (\ref{K_nn'(q_e)=0}) for $\nu'=1,2,3$.
(ii) Next, we set  $\nu'=4$ and $q= q_{\rm e}^{}$ in Eq.\ (\ref{K_nn'(q)-21}),
substitute Eq.\ (\ref{Lambda^(4,2)_4(q_e)}),
and transform the resulting expression as follows:
\begin{align*}
&\,K_{\nu4}^{}(q_{\rm e}^{})
\notag \\
=&\, - \frac{1}{2\beta}\sum_{p}\vec\lambda_\nu^{\,{\rm T}}(-{\bf p})\underline{GG}(p;q_{\rm e}^{})\Biggl[
\frac{\hat{G}^{-1}(p_{+{\rm e}}^{})\hat\sigma_3^{}+\hat\sigma_3^{}\hat{G}^{-1}(p)}{i\omega_\ell} 
\notag \\
&\,-\frac{\underline{\Gamma}^{(3){\rm T}}(p;q_{\rm e}^{})\hat\sigma_3^{}\vec{\Psi}}{i\omega_\ell}\Biggr] + V\bigl[\hat\Lambda_\nu^{(2)}(-q_{\rm e}^{})\vec\Psi\bigr]^{\rm T} \frac{\hat\sigma_3^{}\vec\Psi }{i\omega_\ell^{}}
\notag \\
=&\, \frac{1}{2\beta}\!\sum_{p}{\rm Tr}\,\hat\lambda_\nu^{}({\bf p})
\frac{\hat\sigma_3^{}\hat{G}(p)+\hat{G}(p_{+{\rm e}}^{})\hat\sigma_3^{}}{i\omega_\ell} 
+ V\vec\Psi^{\rm T}\hat\lambda_\nu^{\rm T}({\bf 0})
\frac{\hat\sigma_3^{}\vec\Psi }{i\omega_\ell^{}} ,
\end{align*}
where we have used Eqs.\ (\ref{GGB}) and (\ref{Lambda^(2)T}).
The last expression can be shown to vanish for each of $\nu=1,2,3$ and $\nu=4$ by substituting Eq.\ (\ref{lambda_nu})
and making a change of variables $p_{+{\rm e}}^{}\rightarrow p$ in the summation over $\hat{G}(p_{+{\rm e}}^{})$.
Thus, we arrive at Eq.\ (\ref{K_nn'(q_e)=0}) for $\nu'=4$.
(iii) Finally, the symmetry of Eq.\ (\ref{K_nn'-symm}) completes the proof of Eq.\ (\ref{K_nn'(q_e)=0}).

Third, $K_{\nu\nu'}^{}(q)$ for $q=q_{\rm m}^{}\equiv ({\bf q},0)$ and
$\nu,\nu'=1,2,3$ satisfy the {\it  longitudinal sum rule} or {\it f-sum rule} \cite{PN66} given by
\begin{align}
\sum_{\nu'=1}^3 K_{\nu\nu'}^{}(q_{\rm m}^{})q_{\nu'}^{}=-\frac{N}{m}q_{\nu}^{} ,
\label{Lsum}
\end{align}
where $N$ is the number of particles in the system.
The identity is proved as follows. We multiply Eq.\ (\ref{K_nn'(q)-21})  for $q=q_{\rm m}^{}$ by $q_{\nu'}^{}$ and
take the sum over $\nu'=1,2,3$, substitute Eq.\ (\ref{Lambda^(4,2)(q_m)}), and transform the resulting equation as
follows:
\begin{align*}
&\,\sum_{\nu'=1}^3 K_{\nu\nu'}^{}(q_{\rm m}^{})q_{\nu'}^{}
\notag \\
=&\,
- \frac{1}{2\beta}\sum_{p}\vec\lambda_\nu^{\,{\rm T}}(-{\bf p})\underline{GG}(p;q_{\rm m}^{})\Bigl[
\hat{G}^{-1}(p_{+{\rm m}}^{})\hat\sigma_3^{}
\notag \\
&\,+\hat\sigma_3^{}\hat{G}^{-1}(p_{-{\rm m}}^{})-\underline{\Gamma}^{(3){\rm T}}(p;q_{\rm m}^{})\hat\sigma_3^{}\vec{\Psi}\Bigr] 
\notag \\
&\,+ V\bigl[\hat\Lambda_\nu^{(2)}(-q_{\rm m}^{})\vec\Psi\bigr]^{\rm T} \hat\sigma_3^{}\vec\Psi
\notag \\
=&\, - \frac{1}{2\beta}\sum_{p}{\rm Tr}\,\hat\lambda_\nu^{\,{\rm T}}({\bf p})\Bigl[
\hat\sigma_3^{}\hat{G}(p_{-{\rm m}}^{})
+\hat{G}(p_{+{\rm m}})\hat\sigma_3^{}\Bigr] 
\notag \\
&\,+ V\vec\Psi^{\rm T}\hat\lambda_\nu({\bf q}/2) \hat\sigma_3^{}\vec\Psi
\notag \\
=&\, -\left[- \frac{1}{2\beta} \sum_{p}{\rm Tr}\,\hat\sigma_1^{}\hat{G}(p)
+\frac{V}{2}\vec\Psi^{\rm T}\hat\sigma_1^{}\vec\Psi\right]\frac{q_\nu^{}}{m}
\notag \\
=&\, -\frac{N}{m}q_\nu^{} ,
\end{align*}
where we have used Eqs.\ (\ref{GGB}) and (\ref{Lambda^(2)T}), substituted $\lambda_\nu^{}({\bf p})\!=\!(p_\nu^{}/m)i\hat\sigma_2^{}$ from Eq.\ (\ref{lambda_nu}), made a change of variables $p_{\pm{\rm m}}^{}\!\rightarrow\! p$ for the sum over $\hat{G}(p_{\pm{\rm m}}^{})$,  
noted that the sum of ${\bf p}\,\hat{G}(p)$ over ${\bf p}$ vanishes, 
and replaced $G(p)$ by $G( p)e^{i\varepsilon_n 0^+}$ in the final sum over $p$
following the standard procedure for the equal-time Green's function \cite{LW60}.

Fourth, the current correlation functions ($\nu,\nu'=1,2,3$) of the normal state ($\vec\Psi=\vec0$) in the $q$-limit satisfy
\begin{align}
K_{\nu\nu'}^{\rm n}(q_{\rm m}^{}\rightarrow 0)=-\frac{N}{m}\delta_{\nu\nu'}^{} ,
\label{K_ii'^n}
\end{align}
which implies that the normal density is equal to the particle density in the normal state, as it should.
The proof proceeds as follows. We take the $q$-limit of Eq.\ (\ref{K_nn'(q)-21}),
substitute Eq.\ (\ref{Lambda^(4)(q_m)-2}), and set $\vec\Psi=\vec0$. The resulting expression can be transformed 
by using Eq.\ (\ref{GGB}) and noting $\partial\hat{G}^{-1}(p_{\pm {\rm m}})/\partial q_\nu^{}\bigr|_{{\bf q}={\bf 0}}=\pm \frac{1}{2}\partial\hat{G}^{-1}(p)/\partial p_\nu^{}$ as
\begin{align*}
&\,K_{\nu\nu'}^{\rm n}(q_{\rm m}^{}\rightarrow 0)
\notag \\
=&\, - \frac{1}{2\beta}\sum_{p}\vec\lambda_\nu^{\,{\rm T}}(-{\bf p})\underline{GG}^q(p)\vec\Lambda^{(4q)}_\nu(p)
\notag \\
=&\, - \frac{1}{4\beta}\sum_{p}\hat\lambda_\nu({\bf p})\hat{G}(p)
\!\left[\frac{\partial\hat{G}^{-1}(p)}{\partial p_{\nu'}}
\hat\sigma_3^{}-\hat\sigma_3^{}\frac{\partial\hat{G}^{-1}(p)}{\partial p_{\nu'}}\right]\!\hat{G}(p)
\notag \\
=&\,  -\frac{1}{4\beta}\sum_{p}{\rm Tr} \,\hat\lambda_\nu^{}({\bf p})
\left[\frac{\partial\hat{G}(p)}{\partial p_{\nu'}}\hat\sigma_3^{}-\hat\sigma_3^{}\frac{\partial\hat{G}(p)}{\partial p_{\nu'}}\right]
\notag \\
=&\, \frac{1}{2\beta}\sum_{p}\frac{\delta_{\nu\nu'}}{m}{\rm Tr} \,
\hat\sigma_1^{}\hat{G}(p)
\notag \\
=&\, -\delta_{\nu\nu'}\frac{N}{m},
\end{align*}
where we have (i) also used $\hat{G}\hat\sigma_3^{}\!=\!-\hat\sigma_3^{}\hat{G}$ that holds for the normal state and $\hat{G}\,\delta\hat{G}^{-1}\hat{G}=-\delta\hat{G}$, 
(ii) substituted $\hat\lambda_\nu^{}({\bf p})=(p_\nu^{}/m)i\hat\sigma_2^{}$ from Eq.\ (\ref{lambda_nu}), 
and (iii) performed integration by parts with respect to ${\bf p}$.
Thus, we have derived Eq.\ (\ref{K_ii'^n}) for the normal state.

Fifth,  the density correlation function $K_{44}(q)$ in the $q$-limit satisfies the {\it  compressibility sum rule} \cite{PN66},
\begin{align}
K_{44}^q =-\frac{\partial N}{\partial\mu} .
\label{CSR}
\end{align}
To prove it, we set $\nu=\nu'=4$ in Eq.\ (\ref{K_nn'(q)-21}), take the $q$-limit 
and express the second term on the right-hand side in terms of Eq.\ (\ref{dPsiG/dmu11}), 
substitute Eqs.\ (\ref{lambda_nu}) and (\ref{Lambda^(4q,2q)_4}), 
and transform the resulting expression as follows
\begin{align*}
K_{44}^q
=&\, - \frac{1}{2\beta}\sum_{p}\vec\sigma_1\underline{GG}^q(p)\vec\Lambda_4^{\,(4q)}(p)
-V\bigl(\hat\Lambda^{(2q)}_4\vec\Psi)^{\rm T}\frac{\partial\vec\Psi}{\partial\mu}
\notag \\
=&\, - \frac{1}{2\beta}\sum_{p}\vec\sigma_1\underline{GG}^q(p)\!
\left[\frac{\partial \vec{G}^{-1}(p)}{\partial\mu}+\underline{\Gamma}^{(3q){\rm T}}(p)\frac{\partial\vec{\Psi}}{\partial\mu}\right]
\notag \\
=&\, \frac{1}{2\beta}\sum_{p}{\rm Tr}\,\hat\sigma_1^{}
\frac{\partial \hat{G}(p)}{\partial\mu}-V\vec{\Psi}^{\,{\rm T}}\hat\sigma_1^{}\frac{\partial\vec{\Psi}}{\partial\mu}
\notag \\
=&\, -\frac{\partial N}{\partial\mu},
\end{align*}
where we have also used Eq.\ (\ref{GGB}), $\hat{G}\,\delta\hat{G}^{-1}\hat{G}=-\delta\hat{G}$, and Eq.\ (\ref{identity-q-limit}).

\subsection{Correlation functions in terms of $\underline{\Gamma}^{(4\omega)}$}

We finally express the correlation functions in terms of low-energy vertices 
$\underline{\Gamma}^{(4\omega)}$ in the $\omega$-limit.
The four-point vertex $\underline{\Gamma}^{(4)}(p,p';q)$ satisfies
\begin{align}
\underline{\Gamma}^{(4)}(p,p';q)=&\,\underline{\Gamma}^{(4{\rm i})}(p,p';q)-\frac{1}{2\beta V}\sum_{p''}\underline{\Gamma}^{(4{\rm i})}(p,p'';q)
\notag \\
&\,\times \underline{GG}(p'';q)
\underline{\Gamma}^{(4)}(p'',p';q),
\label{BS}
\end{align}
where $\underline{\Gamma}^{(4{\rm i})}(p,p';q)$ is the irreducible four-point vertex; see Eq.\ (58a) of I.
Its solution can be written symbolically with omitting  the factor $(\beta V)^{-1}$ for simplicity as 
\begin{align}
\underline{\Gamma}^{(4)}=\left(\underline{1}+\frac{1}{2}\underline{\Gamma}^{(4{\rm i})}
\underline{GG}\right)^{-1}\underline{\Gamma}^{(4{\rm i})}.
\label{Gamma^(4)-def}
\end{align}
Let us express $\underline{GG}$ as a sum of the two contributions,
\begin{subequations}
\begin{align}
\underline{GG}=\underline{GG}^\omega+\underline{GG}^{\rm L} ,
\label{GG-decompose}
\end{align}
where $\underline{GG}^\omega\equiv \underline{GG}^\omega(p)$ denotes the $\omega$-limit of $\underline{GG}$ defined by
\begin{align}
\underline{GG}^\omega(p)\equiv \lim_{q_{\rm e}^{}\rightarrow 0}\underline{GG}(p;q_{\rm e}^{})
\label{GG^omega}
\end{align}
\end{subequations}
with $q_{\rm e}^{}\equiv ({\bf 0},\omega_\ell^{})$, and superscript $^{\rm L}$ signifies {\it low energy}.
Then $\underline{\Gamma}^{(4)}$ can be written alternatively as
\begin{align}
\underline{\Gamma}^{(4)}=\left(\underline{1}+\frac{1}{2}\underline{\Gamma}^{(4\omega)}
\underline{GG}^{\rm L}\right)^{-1}\underline{\Gamma}^{(4\omega)},
\label{Gamma^(4)-LE}
\end{align}
with $\underline{\Gamma}^{(4\omega)}\equiv\bigl(\underline{1}+\frac{1}{2}\underline{\Gamma}^{(4{\rm i})}
\underline{GG}^\omega\bigr)^{-1}\underline{\Gamma}^{(4{\rm i})}$;
see Appendix\ref{sec:App3} for the derivation.

It is also shown in Appendix\ref{sec:App3} that Eq.\ (\ref{K_nn'(q)-2}) is expressible in terms of $\underline{GG}^{\rm L}$ as
\begin{align}
K_{\nu\nu'}(q)
=&\,K_{\nu\nu'}^{{\rm r}\omega}- \frac{1}{2\beta}\sum_{p}\vec\Lambda_\nu^{\,(4\omega){\rm T}}(-p)\underline{GG}^{\rm L}(p;q)
\vec\Lambda_{\nu'}^{(4)}(p;q) 
\notag \\
&\, + V\bigl[\hat\Lambda_\nu^{(2)}(-q)\vec\Psi\bigr]^{\rm T} \hat{G}(q)\,\hat\Lambda_{\nu'}^{(2)}(q) \vec\Psi ,
\label{K_nn'(q)-3}
\end{align}
where $K_{\nu\nu'}^{{\rm r}\omega}$ denotes
\begin{align}
K_{\nu\nu'}^{{\rm r}\omega}=&\,\delta_{\nu 4}\delta_{\nu' 4}
\biggl\{
-\frac{\partial N}{\partial\mu}
+\frac{1}{2\beta}\sum_p\vec\Lambda_4^{\,(4\omega){\rm T}}(-p) 
\notag \\
&\,\times \bigl[\underline{GG}^{q}(p)-
\underline{GG}^\omega(p)\bigr] \vec\Lambda_4^{(4q)}(p)\biggr\} ,
\end{align}
and $\hat\Lambda_{\nu}^{(2)}(q)\vec\Psi$ and $\vec\Lambda_{\nu}^{(4)}(p;q)$ are given by
\begin{subequations}
\label{Lambda^(2,4)-R1}
\begin{align}
\hat\Lambda_\nu^{\,(2)}(q)\vec\Psi=&\,\hat\Lambda_\nu^{\,(2\omega)}\vec\Psi
-\frac{1}{\beta V}\sum_p\underline{\Gamma}^{(3)}(p;q) \underline{GG}^{\rm L}(p;q)
\notag \\
&\,\times \vec\Lambda_\nu^{\,(4\omega)}(p),
\label{Lambda^(2,4)-R11}
\end{align}
\begin{align}
\vec\Lambda_\nu^{\,(4)}(p;q)=&\,\vec\Lambda_\nu^{\,(4\omega)}(p)-\frac{1}{2\beta V}\sum_{p'} \underline{\Gamma}^{(4)}(p,p';q)\underline{GG}^{\rm L}(p';q)
\notag \\
&\,\times \vec\Lambda_\nu^{\,(4\omega)}(p') .
\label{Lambda^(2,4)-R12}
\end{align}
\end{subequations}
Comparing Eq.\ (\ref{Lambda^(2,4)-R1}) with Eq.\ (\ref{Lambda^(4,2)}), we observe that the bare vertices
$\hat\lambda_\nu\vec\Psi$ and $\vec\lambda_\nu$ have been replaced by
renormalized vertices $\hat\Lambda_{\nu}^{(2\omega)}\vec\Psi$ and 
$\vec\Lambda_{\nu}^{(4\omega)}(p)$ 
in the $\omega$-limit, which are expressible in terms of $\hat{G}^{-1}$ by Eq.\ (\ref{Lambda^(4,2)_4(q_e)}) for $\nu=4$ 
and by Eq.\ (\ref{Lambda^(4,2)_j(q_e)}) for $\nu=1,2,3$ with $q_{\rm e}\rightarrow 0$.

\section{Summary\label{sec:5}}

We have derived four kinds of Ward-Takahashi identities for correlated Bose-Einstein condensates,
which are given by Eqs.\ (\ref{WT-gauge3}), (\ref{WT-gauge4}), (\ref{WT-Galilean-a}), and (\ref{dPsiG/dmu1}).
Each of them consists of the condensate part and the quasiparticle part, and taking the limit of $\Psi\rightarrow 0$ yields
the normal-Bose-liquid versions of the four identities well known in the theory of normal Fermi liquids \cite{AGD63}.
Compared with the two identities on the quasiparticle part obtained by Gavoret and Nozi\`eres at $T=0$ based on the fictitious gap method \cite{GN64},
i.e., their Eqs.\ (5.22) and (5.24), our identities (i) have the condensate part, 
(ii) are distinct between the $\omega$- and $q$-limits, and (iii) contain additional terms 
with the three-point vertices.

We have also obtained expressions of the density and current correlation functions
as Eq.\ (\ref{K_nn'(q)-2}), which are composed of the regular part and singular part
in agreement with the result by Gavoret and Nozi\`eres. 
Using the Ward-Takahashi identities, we have derived exact properties of the correlation functions,
which are given by Eqs.\ (\ref{K_nn'-symm})-(\ref{CSR}).
They include the well-known longitudinal and compressibility sum rules.
It follows from Eqs.\ (\ref{Lsum}) and (\ref{K_ii'^n}) 
that the finite superfluid density emerges due to $\vec\Psi\neq \vec0$ 
along the direction perpendicular to ${\bf q}$.
The fact that the vertices are different between the $\omega$-limit and $q$-limit strongly suggests that
the regular part of the correlation functions, i.e., the first term of Eq.\ (\ref{K_nn'(q)-2}), can sustain
collective oscillations in the same way as in Fermi liquids \cite{Landau58,AGD63} and superfluid Fermi liquids \cite{Leggett66,SR83}.

Finally, the correlations functions have been shown to be expressible in terms of low-energy Green's functions and vertices as Eq.\ (\ref{K_nn'(q)-3}). They will form a basis for constructing the low-energy effective theory of 
correlated Bose-Einstein condensates.

Although the present consideration is restricted to equilibrium ones, the extension to nonequilibrium systems
can be performed straightforwardly based on the formulation on the Keldysh contour of I 
by choosing $p_\pm^{}$ symmetrically as $p_\pm^{}\!\equiv \!p\!\pm\!q/2$ instead of those in Eq.\ (\ref{p_pm}).

\section*{Acknowledgment}
This work was supported by JSPS KAKENHI Grant Number JP20K03848.

\appendix

\section{Symmetry Properties of Eq.\ (\ref{VecMat(q)})\label{sec:App1}}

It follows from Eq.\ (6b)  of I
that Green's functions in the coordinate space obey $G_{jj'}(x,x')=G_{j'j}(x',x)$. Moreover, they satisfy \cite{Kita19-1} $G_{jj'}(x,x')=\bigl[G_{3-j',3-j}({\bf r}'\tau,{\bf r}\tau')\bigr]^*$.
The relations translate through Eq.\ (\ref{G-exp}) into
\begin{align}
G_{jj'}(p)=G_{j'j}(-p)=&\,\bigl[G_{3-j',3-j}({\bf p},-\varepsilon_n)\bigr]^*
\notag \\
=&\,G_{3-j',3-j}(p) .
\label{G-symm}
\end{align}
The last equality holds in the gauge where the condensate wave function is real.
Using Eq.\ (\ref{G-symm}), one can show easily that Eq.\ (\ref{GG(p;q)}) satisfies
\begin{align}
\langle j_1j_1'| \underline{GG}(p;q) |j_2j_2'\rangle
=\langle j_2j_2'| \underline{GG}(-p;-q) |j_1j_1'\rangle .
\label{GG-symm}
\end{align}

It also follows from Eqs.\ (28a) and (35a) of I that the four-point vertices obey
\begin{align*}
\Gamma^{(4)}(\xi_1,\xi_1';\xi_2,\xi_2')=&\,\Gamma^{(4)}(\xi_1',\xi_1;\xi_2,\xi_2')
\notag \\
=&\,\Gamma^{(4)}(\xi_2,\xi_2';\xi_1,\xi_1') .
\end{align*}
Accordingly, the Fourier coefficients in Eq.\ (\ref{Gamma^(4)-exp}) satisfy
\begin{align}
\Gamma^{(4)}_{j_1j_1';j_2j_2'}(p,p';q) =&\,\Gamma^{(4)}_{j_1'j_1;j_2j_2'}(-p_{+{\rm e}}^{},p';q)
\notag \\
=&\, \Gamma^{(4)}_{j_2j_2';j_1j_1'}(-p',-p;-q) ,
\label{Gamma^(4)-symm}
\end{align}
with $p_{+{\rm e}}^{}\equiv ({\bf p},\varepsilon_n+\omega_\ell^{})$.

Regarding $\underline{\Gamma}^{(3)}(p;q)$, one can show based on Eq.\ (35b) of I that it is connected
with $\underline{\Gamma}^{(4)}(p,p';q)$ by 
\begin{align}
\Gamma_{j_1;j_2j_2'}^{(3)}(p;q)=\sum_{j_1'}(-1)^{j_1+j_1'-1}\Psi_{j_1'}\Gamma_{j_1j_1';j_2j_2'}^{(4)}(q_{\rm m}^{}/2,p;q),
\label{Gamma^(3)-Gamma^(4)}
\end{align}
with $q_{\rm m}^{}\!\equiv\!({\bf q},0)$, which
obeys
\begin{align}
\Gamma^{(3)}_{j_1;j_2j_2'}(p;q) =\Gamma^{(3)}_{j_1;j_2'j_2}(-p_{+{\rm e}}^{};q)
=\Gamma^{(3){\rm T}}_{j_2j_2';j_1}(-p;-q) ,
\label{Gamma^(3)-Gamma^(3)T}
\end{align}
as shown by using Eqs.\ (35b) and (35c) of I and Eq.\ (\ref{Gamma^(4)-symm}) above.

\section{Derivation of Eq.\ (\ref{identity-q-limit})\label{sec:App2}}

We here derive Eq.\ (\ref{identity-q-limit}) from Eq.\ (\ref{dPsiG^-1/dmu12}).
Let us multiply Eq.\ (\ref{dPsiG^-1/dmu12}) at $p\!=\!0$ by $-\hat\sigma_3^{}\underline{\Psi}^{(3\sigma)}$ 
from the left, where $\underline{\Psi}^{(3\sigma)}$ is defined by
\begin{align}
\langle j_1|\underline{\Psi}^{(3\sigma)}|j_2j_2'\rangle \equiv \delta_{j_1j_2}^{}(-1)^{j_2'-1}\Psi_{j_2'}^{}. 
\end{align}
The left-hand side of the resulting equation yields
\begin{align}
-\hat\sigma_3^{}\underline{\Psi}^{(3\sigma)}\frac{\partial \vec{G}^{-1}(0)}{\partial\mu}
=-\hat\sigma_3^{}\frac{\partial \hat{G}^{-1}(0)}{\partial\mu}\hat\sigma_3^{}\vec{\Psi} =\vec{0},
\label{AppB-left}
\end{align}
where we have (i) made a transformation similar to Eq.\ (42) of I
and also (ii) used the Hugenholtz-Pines relation \cite{HP59} $\hat{G}^{-1}(0)\hat\sigma_3^{}\vec{\Psi}/\Psi=0$ in the gauge $\Psi_1=\Psi_2\equiv \Psi$.
On the other hand, the three terms on the right-hand side are transformed as
\begin{subequations}
\label{AppB-right}
\begin{align}
-\hat\sigma_3^{}\underline{\Psi}^{(3\sigma)}\vec\sigma_1=-\hat\sigma_3^{}\hat\sigma_1^{}\hat\sigma_3^{}\vec{\Psi}
=\hat\sigma_1^{}\vec{\Psi},
\label{AppB-right1}
\end{align}
\begin{align}
&\,-\frac{1}{2\beta V}\sum_{p'}\bigl(-\hat\sigma_3^{}\underline{\Psi}^{(3\sigma)}\bigr) \underline{\Gamma}^{(4q)}(0,p')\underline{GG}^q(p')\vec\sigma_1
\notag \\
=&\, -\frac{1}{2\beta V}\sum_{p'} \underline{\Gamma}^{(3q)}(p')\underline{GG}^q(p')\vec\sigma_1,
\end{align}
\begin{align}
-\bigl(-\hat\sigma_3^{}\underline{\Psi}^{(3\sigma)}\bigr)\underline{\Gamma}^{(3q){\rm T}}(0)\frac{\partial\vec{\Psi}}{\partial\mu}=-\left[\hat{\Sigma}(0)+\hat\sigma_3^{}\hat{\Sigma}(0)\hat\sigma_3^{}\right]
\frac{\partial\vec{\Psi}}{\partial\mu} ,
\label{AppB-right3}
\end{align}
\end{subequations}
where we have used the Fourier transforms of Eqs.\ (35b) and (19) in I; 
see also Eq.\ (40) of I for the transformation of Eq.\ (\ref{AppB-right3}).
Let us collect Eqs.\ (\ref{AppB-right1})-(\ref{AppB-right3}), equate it with Eq.\ (\ref{AppB-left}),  and use the fact
that Eq.\ (\ref{AppB-right3}) vanishes due to the Nepomnyashchi\u{i} identity \cite{Nepomnyashchii75} $\Sigma_{jj}(0)=0$.
We thereby obtain Eq.\ (\ref{identity-q-limit}).

\section{Derivations of Eqs.\ (\ref{Gamma^(4)-LE})\label{sec:App3}}

It follows from Eq.\ (\ref{BS}) that $\underline{\Gamma}^{(4)}$, $\underline{\Gamma}^{(4{\rm i})}$, and
\begin{align}
\underline{g}\equiv\frac{1}{2}\underline{GG}
\end{align}
obey
$\underline{\Gamma}^{(4)}=\underline{\Gamma}^{(4{\rm i})}-\underline{\Gamma}^{(4{\rm i})}\underline{g}\,\underline{\Gamma}^{(4)}$;
we omit factors $\beta^{-1}$ and $V^{\pm 1}$.
Its formal solution is given by 
\begin{align}
\underline{\Gamma}^{(4)}=&\,\left(\underline{1}+\underline{\Gamma}^{(4{\rm i})}\underline{g}\right)^{-1}\underline{\Gamma}^{(4{\rm i})}
\notag \\
=&\,\underline{\Gamma}^{(4{\rm i})}\left(\underline{1}+\underline{g}\,\underline{\Gamma}^{(4{\rm i})}\right)^{-1} .
\label{Gamma^(4)-defA}
\end{align}
Let us decompose $\underline{g}$ into two parts
\begin{align}
\underline{g}=\underline{g}^{\omega}+\underline{g}^{{\rm L}},
\label{g-decomp}
\end{align}
where $\underline{g}^{\omega}$ is the $\omega$-limit of $\underline{g}$ defined by Eq.\ (\ref{GG^omega}).
We also introduce $\underline{\Gamma}^{(4\omega)}$ by
\begin{align}
\underline{\Gamma}^{(4\omega)}=\left(\underline{1}+\underline{\Gamma}^{(4{\rm i})}\underline{g}^\omega\right)^{-1}\underline{\Gamma}^{(4{\rm i})} .
\label{Gamma^4omega-def}
\end{align}
Then $\underline{\Gamma}^{(4)}$ is expressible 
in terms of $\underline{\Gamma}^{(4\omega)}$ and $\underline{g}^{\rm L}$ as \cite{Landau58,Leggett65}
\begin{align}
\underline{\Gamma}^{(4)}=\left(\underline{1}+\underline{\Gamma}^{(4\omega)}\underline{g}^{\rm L}\right)^{-1}\underline{\Gamma}^{(4\omega)} .
\label{Gamma^(4)-g^L}
\end{align}
This can be shown from  Eq.\ (\ref{Gamma^(4)-defA}) by writing $\underline{1}+\underline{\Gamma}^{(4{\rm i})}\underline{g}=\underline{A}+\underline{B}$ with $\underline{A}\equiv \underline{1}+\underline{\Gamma}^{(4{\rm i})}
\underline{g}^\omega$ and $\underline{B}\equiv \underline{\Gamma}^{(4{\rm i})}
\underline{g}^{\rm L}$,
and using the matrix identity $(\underline{A}+\underline{B})^{-1}=(\underline{1}+\underline{A}^{-1}\underline{B})^{-1}\underline{A}^{-1}$.

Next, $\underline{\Gamma}^{(4)}\underline{g}$,
 $\underline{1}-\underline{\Gamma}^{(4)}\underline{g}$, and $\underline{g}(\underline{1}-\underline{\Gamma}^{(4)}\underline{g})$ can be transformed into \cite{Leggett65}
\begin{subequations}
\label{Gamma^(4)g}
\begin{align}
\underline{\Gamma}^{(4)}\underline{g}=&\, \underline{\Gamma}^{(4\omega)}\underline{g}^\omega
+\underline{\Gamma}^{(4)}\underline{g}^{\rm L}\underline{R},
\label{Gamma^(4)g1}
\\
\underline{1}-\underline{\Gamma}^{(4)}\underline{g}=&\, 
\bigl(\underline{1}-\underline{\Gamma}^{(4)}\underline{g}^{\rm L}\bigr)\underline{R},
\label{Gamma^(4)g2}
\\
\underline{g}(\underline{1}-\underline{\Gamma}^{(4)}\underline{g})=&\, \underline{g}^\omega\underline{R}
+\underline{R}^{\rm T}\underline{g}^{\rm L}(\underline{1}-\underline{\Gamma}^{(4)}\underline{g}^{\rm L})\underline{R} ,
\label{Gamma^(4)g3}
\end{align}
\end{subequations}
with
\begin{subequations}
\label{R}
\begin{align}
\underline{R}\equiv&\, \underline{1}-\underline{\Gamma}^{(4\omega)}\underline{g}^\omega,
\label{R1}
\\
 \underline{R}^{\rm T}\equiv&\, \underline{1}-\underline{g}^\omega\underline{\Gamma}^{(4\omega)}.
\label{R2}
\end{align}
\end{subequations}
Equation (\ref{Gamma^(4)g1}) can be proved by substituting Eq.\ (\ref{g-decomp}) into the left-hand side,
expressing $\underline{\Gamma}^{(4)}\underline{g}^\omega=\bigl(\underline{\Gamma}^{(4\omega)}
-\underline{\Gamma}^{(4)}\underline{g}^{\rm L}\underline{\Gamma}^{(4\omega)}\bigr)\underline{g}^\omega$
based on Eq.\ (\ref{Gamma^(4)-g^L}),
and collecting terms with $\underline{\Gamma}^{(4)}\underline{g}^{\rm L}$.
Equation (\ref{Gamma^(4)g2}) results directly from Eqs.\  (\ref{Gamma^(4)g1}) and (\ref{R1}).
Finally, proof of Eq.\ (\ref{Gamma^(4)g3}) proceeds by writing 
$\underline{g}(\underline{1}-\underline{\Gamma}^{(4)}\underline{g})
=\underline{g}\,\underline{R}-\underline{g}\,\underline{\Gamma}^{(4)}\underline{g}^{\rm L}\underline{R}$ based on Eq.\ (\ref{Gamma^(4)g2}),
expressing $\underline{g}\,\underline{\Gamma}^{(4)}=\underline{g}^\omega\underline{\Gamma}^{(4\omega)}
+\underline{R}^{\rm T}\underline{g}^{\rm L}\underline{\Gamma}^{(4)}$ similarly as Eq.\ (\ref{Gamma^(4)g1}),
and collecting terms with $\underline{g}^{\rm L}$.

We can also write $\underline{\Gamma}^{(3)}\underline{g}$ and $\underline{g}\,\underline{\Gamma}^{(3){\rm T}}$ as
\begin{subequations}
\label{Gamma^(3)g}
\begin{align}
\underline{\Gamma}^{(3)}\underline{g}=&\,\underline{\Gamma}^{(3\omega)}\underline{g}^\omega
+\underline{\Gamma}^{(3)}\underline{g}^{\rm L}\underline{R},
\label{Gamma^(3)g1}
\\
\underline{g}\,\underline{\Gamma}^{(3){\rm T}}=&\,\underline{g}^\omega\underline{\Gamma}^{(3\omega){\rm T}}
+\underline{R}^{\rm T}\underline{g}^{\rm L}\underline{\Gamma}^{(3){\rm T}},
\label{Gamma^(3)g2}
\end{align}
\end{subequations}
as shown by using Eqs.\ (\ref{Gamma^(3)-Gamma^(4)}), (\ref{Gamma^(3)-Gamma^(3)T}),  and (\ref{Gamma^(4)g1}).

Substitution of Eqs.\ (\ref{Gamma^(4)g}) and (\ref{Gamma^(3)g}) into Eq.\ (\ref{K_nn'(q)-2}) yields
\begin{align}
K_{\nu\nu'}=&\, K_{\nu\nu'}^{{\rm r}\omega}-\vec\lambda_\nu^{{\rm T}}\underline{R}^{\rm T}\underline{g}^{\rm L}_{}\vec\Lambda_\nu^{\,(4)}
+ \bigl(\hat\Lambda_\nu^{\,(2)} \vec\Psi\bigr)^{\rm T}\hat{G}\hat\Lambda_\nu^{\,(2)} \vec\Psi,
\label{K_nn'-4}
\end{align}
where $K_{\nu\nu'}^{{\rm r}\omega}$ ($^{\rm r}$ denoting ``regular'') is defined by 
\begin{align}
K_{\nu\nu'}^{{\rm r}\omega}\equiv &\,-\vec\lambda_\nu^{\,{\rm T}}\underline{g}^\omega
\underline{R}\,\vec\lambda_\nu,
\label{K_nn'^romega}
\end{align}
and $\hat\Lambda_\nu^{\,(2)}\vec\Psi$ and $\vec\Lambda_\nu^{\,(4)}$ are now given in terms of 
$\underline{g}^{\rm L}$ and $\underline{R}$ by
\begin{subequations}
\label{Lambda^(2,4)-R2}
\begin{align}
\hat\Lambda_\nu^{\,(2)}\vec\Psi=\hat\lambda_\nu\vec\Psi
-\underline{\Gamma}^{(3\omega)} \underline{g}^\omega\vec\lambda_\nu
-\underline{\Gamma}^{(3)} \underline{g}^{\rm L}\underline{R}\,\vec\lambda_\nu ,
\label{Lambda^(2,4)-R21}
\end{align}
\begin{align}
\vec\Lambda_\nu^{\,(4)}=&\,\left(\underline{1}-\underline{\Gamma}^{(4)}\underline{g}^{\rm L}\right)\underline{R}\,\vec\lambda_\nu .
\label{Lambda^(2,4)-R22}
\end{align}
\end{subequations}

The renormalized vertex $\underline{R}\,\vec\lambda_\nu$ 
in Eqs.\ (\ref{K_nn'^romega}) and (\ref{Lambda^(2,4)-R2}) can be approximated as
\begin{subequations}
\label{Rlambda}
\begin{align}
\underline{R}(p)\vec\lambda_\nu({\bf p})\approx \vec\Lambda^{(4\omega)}_\nu(p).
\label{Rlambda1}
\end{align}
Indeed, $\underline{R}\vec\lambda_\nu\!\equiv \!(\underline{1}-\underline{\Gamma}^\omega \underline{g}^\omega)\vec\lambda_\nu$
differs from Eq.\ (\ref{Lambda^(4)}) in the $\omega$-limit in that 
the argument of $\Gamma^{(4{\rm i})}$ in Eq.\ (\ref{Gamma^4omega-def})
is $q$ instead of $q_{\rm e}^{}(\rightarrow 0)$. Hence,
Eq.\ (\ref{Rlambda1})  is exact in the static homogeneous limit, 
and also expected to hold true to an excellent approximation for any small $q$.
Similarly, the first two terms on the right-hand side 
of Eq.\ (\ref{Lambda^(2,4)-R21}) is expressible
as the $\omega$-limit of Eq.\ (\ref{Lambda^(2)}), i.e.,
\begin{align}
\hat\lambda_\nu\vec\Psi
-\underline{\Gamma}^{(3\omega)} \underline{g}^\omega\vec\lambda_\nu
\approx\hat\Lambda_\nu^{(2\omega)}\vec\Psi.
\label{Rlambda2}
\end{align}
\end{subequations}
Note that Eq.\ (\ref{Rlambda}) results directly by taking the $\omega$-limit of Eq.\ (\ref{Lambda^(2,4)-R2})
where $\underline{g}^{{\rm L}\omega}$ vanishes, as seen from Eq.\ (\ref{g-decomp}).

Similarly, taking the $\omega$-limit of Eq.\ (\ref{K_nn'-4}) yields
\begin{align*}
K_{\nu\nu'}^{{\rm r}\omega}\approx&\, 
K_{\nu\nu'}^{\omega}-\bigl(\hat\Lambda^{(2\omega)}_\nu\vec\Psi\bigr)\hat{G}^\omega\hat\Lambda^{(2\omega)}_{\nu'}\vec\Psi.
\end{align*}
within the approximation of omitting the $q$ dependence in $\underline{R}$.
Noting Eqs.\ (\ref{Lambda^(2)_j(q_e)}) and (\ref{K_nn'(q_e)=0}), we can conclude that
\begin{subequations}
\label{K_nn'^romega1}
\begin{align}
K_{\nu\nu'}^{{\rm r}\omega}&\,=0
\label{K_nn'^romega11}
\end{align}
holds except for $\nu=\nu'=4$; see also Eq.\ (\ref{WT-Galilean1}).
On the other hand, Eq.\ (\ref{K_nn'^romega}) for $\nu=\nu'=4$ can be transformed as
\begin{align}
K_{44}^{{\rm r}\omega}=&\,K_{44}^q-(K_{44}^q-
K_{44}^{{\rm r}\omega})
\notag \\
\approx&\,-\frac{\partial N}{\partial\mu}+\vec\lambda_4^{{\rm T}}\underline{R}^{\rm T}\underline{g}^{{\rm L}q}_{}\vec\Lambda_4^{\,(4q)}
- \bigl(\hat\Lambda_4^{\,(2q)} \vec\Psi\bigr)^{\rm T}\hat{G}^q\hat\Lambda_4^{\,(2q)} \vec\Psi
\notag \\
=&\, -\frac{\partial N}{\partial\mu}
+\vec\Lambda_4^{\,(4\omega){\rm T}}(\underline{g}^{q}-
\underline{g}^\omega) \vec\Lambda_4^{(4q)},
\label{K_nn'^romega12}
\end{align}
\end{subequations}
where we have successively used Eq.\ (\ref{CSR}), the $q$-limit of Eq.\ (\ref{K_nn'-4}), Eqs.\ (\ref{dPsiG/dmu11}), (\ref{Lambda^(2q)_4}),
(\ref{Rlambda1}), and finally Eq.\ (\ref{g-decomp}) in the $q$-limit.

Let us substitute Eqs.\ (\ref{Lambda^(2,4)-R2})-(\ref{K_nn'^romega1}) into Eq.\ (\ref{K_nn'-4})
and write $\vec\lambda_\nu^{{\rm T}}(-{\bf p})\underline{R}^{\rm T}(p)\approx\vec\Lambda_\nu^{(4\omega){\rm T}}(-p)$
based on the $\omega$-limit of Eq.\ (\ref{Lambda^(4)T}).
We thereby obtain Eq.\ (\ref{K_nn'(q)-3}).


\begin{thebibliography}{99}

\bibitem{Ward50}J. C. Ward, Phys. Rev. {\bf 78}, 182 (1950).
\bibitem{Takanashi57}Y. Takahashi, Nuovo Cimento {\bf 6}, 370 (1957). 
\bibitem{Landau56}L. D. Landau, Zh. Eksp. Teor. Fiz. {\bf 30}, 1058 (1956) [Sov. Phys. JETP {\bf 3}, 920 (1957).
\bibitem{Landau57}L. D. Landau, Zh. Eksp. Teor. Fiz. {\bf 32}, 59 (1957) [Sov. Phys. JETP {\bf 5}, 101 (1957).
\bibitem{Landau58}L. D. Landau, Zh. Eksp. Teor. Fiz. {\bf 35}, 97 (1958) [Sov. Phys. JETP {\bf 8}, 70 (1959).
\bibitem{Pitaevskii59}L. P. Pitaevski\u{i}, Zh. Eksp. Teor. Fiz. {\bf 37}, 1794  (1959) [Sov. Phys. JETP {\bf 10}, 1267 (1960)].
\bibitem{AGD63}A. A. Abrikosov, L. P. Gor'kov, and I. M. Dzyaloshinski, {\it Methods of Quantum Field Theory in Statistical Physics} (Dover, New York, 1975).
\bibitem{NL62}P. Nozi\`eres and J. M. Luttinger, Phys. Rev. {\bf 127}, 1423 (1962).
\bibitem{Leggett65}A. J. Leggett, Phys. Rev. {\bf 140}, A1869 (1965).
\bibitem{Leggett66}A. J. Leggett, Phys. Rev. {\bf 147}, 119 (1966).
\bibitem{SR83}J. W. Serene and D. Rainer, Phys. Rep. {\bf 101}, 221 (1983).
\bibitem{GN64}J. Gavoret and P. Nozi\`eres, Ann. Phys. {\bf 28}, 349 (1964).
\bibitem{Kita20-1}T. Kita, J. Phys. Soc. Jpn. {\bf 90}, 024001  (2021).
\bibitem{Baym69}G. Baym, in {\it Mathemetical Methods in Solid State and Superfluid Theory} (Oliver and Boyd, Edinburgh, 1969), 
ed. R. C. Clark and G. H. Derrick, p. 121.
\bibitem{HB07}M. Holtzmann and G. Baym, Phys. Rev. B {\bf 76}, 092502 (2007).
\bibitem{comment}The sum rule and inequality are given by Eqs.\ [{\sl184\,}] and [{\sl198\,}] in Ref.\ \onlinecite{Baym69}, respectively.
They are obtained based on Eqs.\ [{\sl176\,}], [{\sl177\,}], and [{\sl180\,}],
which in turn have been derived as responses to the perturbation of Eq.\ [{\sl173\,}] by assuming that 
it  affects only the condensate wave function as Eq.\ [{\sl170\,}].
\bibitem{Watabe20}S. Watabe, New J. Phys. {\bf 22}, 103010 (2020).
\bibitem{Kita19-1}T. Kita, J. Phys. Soc. Jpn. {\bf 88}, 054003  (2019).
\bibitem{Kita19-2}T. Kita, J. Phys. Soc. Jpn. {\bf 88}, 104003  (2019).
\bibitem{PN66}D. Pines and P. Nozi\`eres, {\it The Theory of Quantum Liquids Vol. I} (W. A. Benjamin, New York, 1966).
\bibitem{LW60}J. M. Luttinger and J. C. Ward, Phys. Rev. {\bf 118}, 1417 (1960).
\bibitem{HP59}N. M. Hugenholtz and D. Pines, Phys. Rev. {\bf 116}, 489 (1959).
\bibitem{Nepomnyashchii75}A. A. Nepomnyashchi\u{i} and Yu. A. Nepomnyashchi\u{i}, JETP Lett. {\bf 21}, 1 (1975).




\end{thebibliography}
\end{document}